%% file: acl_latex.tex
\pdfoutput=1
\documentclass[11pt]{article}
\usepackage{enumitem}
\usepackage[final]{acl}

\usepackage{times}
\usepackage{latexsym}

\usepackage[T1]{fontenc}
\usepackage[utf8]{inputenc}

\usepackage{microtype}

\usepackage{inconsolata}

\usepackage{graphicx}

\usepackage{comment}
\usepackage{amsthm}
\usepackage{float}
\usepackage{graphicx}
\usepackage{subfigure}
\usepackage{caption}
\usepackage{amsmath,amssymb}
\usepackage{mathtools}
\usepackage{booktabs}
\usepackage{multirow}
\usepackage{siunitx}
\usepackage{adjustbox}
\usepackage{pifont}
\usepackage{scalerel}
\usepackage{tabularray}
\usepackage{makecell}
\usepackage{wrapfig}
\usepackage{enumitem}
\usepackage[most]{tcolorbox}

\definecolor{stepcolor}{HTML}{d79b00}
\definecolor{contentcolor}{HTML}{3439a2}

\usepackage{tikz}

\usepackage{xspace}

\definecolor{stepcolor}{HTML}{d79b00}
\definecolor{contentcolor}{HTML}{6c8ebf}

\usepackage{amssymb}
\usepackage{amsmath} 
\usepackage{url}
\usepackage[utf8]{inputenc} % allow utf-8 input
\usepackage[T1]{fontenc}    % use 8-bit T1 fonts
\usepackage{hyperref}       % hyperlinks
\usepackage{url}            % simple URL typesetting
\usepackage{booktabs}       % professional-quality tables
\usepackage{amsfonts}       % blackboard math symbols
\usepackage{nicefrac}       % compact symbols for 1/2, etc.
\usepackage{microtype}      % microtypography
\usepackage{bm}
\usepackage{graphicx}
\usepackage{times}
\usepackage{latexsym}
\usepackage{mathtools}
\usepackage{amsmath}
\usepackage{pgfplots} 
\usepackage{booktabs}
\usepackage{xspace}
\usepackage{cleveref}
\usepackage{verbatim} % for multiline comments
\usepackage{amsmath,amsfonts,bm}
\usepackage{arydshln}
\usepackage{enumitem}
\usepackage{wrapfig}
\usepackage{selectp}
\usepackage{algorithm}
\usepackage{algorithmic}
\usepackage{times}
\usepackage{latexsym}
\usepackage{amsfonts}
\usepackage{helvet}
\usepackage{url}
\usepackage{float,color}
\usepackage{times}
\usepackage{float}
\usepackage{pdfpages}
\usepackage[commandnameprefix=always]{changes}
\usepackage{epsfig}
\usepackage{stfloats}
\usepackage{url}
\usepackage{diagbox}
\usepackage{subfigure}
\usepackage{epstopdf}
\usepackage{multicol}
\usepackage{makecell}
\usepackage{amsmath}
\usepackage{longtable}
\usepackage{dsfont,amsfonts,color}
\usepackage{multirow}
\usepackage{booktabs}
\usepackage{colortbl}
\PassOptionsToPackage{table}{xcolor}
\usepackage{xr}
\usepackage{tablefootnote}
\usepackage{tabularx}
\usepackage{listings}
\usepackage{wrapfig}
\usepackage{amsthm}
\usepackage{enumitem}
\usepackage[subfigure]{tocloft}

\usepackage{todonotes} % Insert [disable] to disable all notes:
\usepackage{float}
\usepackage[algo2e, lined, ruled,  linesnumbered, commentsnumbered, longend, noend]{algorithm2e}

\input{macro}
\usepackage{cleveref}
\usepackage{xcolor}
\definecolor{lightgrey}{rgb}{0.83, 0.83, 0.83}

\title{RewardDS: Privacy-Preserving Fine-Tuning for Large Language Models via Reward Driven Data Synthesis}

\author{
     \textbf{Jianwei Wang}\textsuperscript{\rm 1}, 
     \textbf{Chengming Shi}\textsuperscript{\rm 1},
     \textbf{Junyao Yang}\textsuperscript{\rm 1}, 
     \textbf{Haoran Li}\textsuperscript{\rm 2}, 
     \textbf{Qianli Ma} \textsuperscript{\rm 1}, \\
     \textbf{Huiping Zhuang}\textsuperscript{\rm 1}, 
     \textbf{Cen Chen}\textsuperscript{\rm 1},
     \textbf{Ziqian Zeng}\textsuperscript{\rm 1}\thanks{Corresponding author} \\
    \textsuperscript{\rm 1} South China University of Technology, 
    \textsuperscript{\rm 2} HKUST \\
    \texttt{wjwfyu@gmail.com}, \texttt{zqzeng@scut.edu.cn}
}

\begin{document}
\maketitle
\begin{abstract}
The success of large language models (LLMs) has attracted many individuals to fine-tune them for domain-specific tasks by uploading their data. 
However, in sensitive areas like healthcare and finance, privacy concerns often arise.
One promising solution is to generate synthetic data with Differential Privacy (DP) guarantees to replace private data. 
However, these synthetic data contain significant flawed data, which are considered as noise.
Existing solutions typically rely on naive filtering by comparing ROUGE-L scores or embedding similarities, which are ineffective in addressing the noise.
To address this issue, we propose \textit{RewardDS}, a novel privacy-preserving framework that fine-tunes a reward proxy model and uses reward signals to guide the synthetic data generation.
Our \textit{RewardDS} introduces two key modules, Reward Guided Filtering and Self-Optimizing Refinement, to both filter and refine the synthetic data, effectively mitigating the noise. 
Extensive experiments across medical, financial, and code generation domains demonstrate the effectiveness of our method.
Our code and data will be available at \url{https://github.com/wjw136/RewardDS}.

\end{abstract}

\input{introduction}

\input{related_work}
\input{method}

\input{privacy}

\input{experiments}

\input{conclusion}

\newpage
\bibliography{anthology}
\input{appendix}

\end{document}

%% file: macro.tex
\DeclareMathOperator*{\argmax}{arg\,max} % Custom operator definition
\DeclareMathOperator*{\argmin}{arg\,min}

\crefname{section}{\S}{\S}
\crefname{table}{Tb.}{Tbs.}
\crefname{appendix}{App.}{Apps.}
\Crefname{theorem}{Thm.}{Thms.}
\Crefname{proposition}{Prop.}{Props.}
\crefname{algorithm}{Alg.}{Algs.}
\Crefname{assumption}{Asm.}{Asms.}
\crefname{mechanism}{Mech.}{Mechs.}
\Crefname{definition}{Def.}{Def.}

\crefformat{footnote}{#2\footnotemark[#1]#3}
\crefmultiformat{footnote}{#2\footnotemark[#1]#3}%
{\textsuperscript{,}#2\footnotemark[#1]#3}{\textsuperscript{,}#2\footnotemark[#1]#3}{\textsuperscript{,}#2\footnotemark[#1]#3}

\definecolor{mygray}{gray}{0.6}

%% file: introduction.tex
\section{Introduction}
The remarkable capabilities of Large Language Models (LLMs) in general tasks have motivated many individuals and organizations to customize their own LLMs for domain-specific applications, such as medical diagnosis, financial analysis, etc. \citep{wu2023bloomberggpt,chen2023meditron}. 
While domain adaptation through fine-tuning is attractive, high computational costs make local fine-tuning impractical for most users. 
Currently, most LLM service providers \citep{achiam2023gpt,YangQwen224,Doubao2024} offer fine-tuning services, allowing users to customize LLMs for their needs by preparing and uploading their domain-specific data. 
However, these data may contain sensitive information, and directly transferring it to the LLM service provider can lead to significant privacy concerns \citep{zeng2024privacyrestore, Sahar2023Not}. 
We denote the individuals or organizations which aim to customize LLMs as \textbf{client}, the LLM service providers as \textbf{server}, and the model being fine-tuned as the \textbf{target LLM}.
It remains a critical challenge to develop privacy-preserving fine-tuning methods in such a client-server scenario.

%However, the collection of this data often requires substantial intellectual and financial investment, making the direct uploading vulnerable to data leakage. 
% }

% 第二段：合成数据是解决这个问题的一个主流的方法。DP是很流行的解决了的问题的framework，DP需要依赖于合成数据。合成数据的范式是什么，合成数据怎么做。在哪合成，怎么合成。不考虑过滤的点，合成数据是怎么做的。

% Following the principles of Differential Privacy(DP) \citep{DBLP:journals/fttcs/DworkR14}, differentially private synthetic text represents a promising and extensively \revisezq{researched approach.} 
% It seeks to generate a new text dataset that retains the characteristics of the original private data \revisezq{while ensuring privacy by safeguarding sensitive information in each sample.} 

\begin{figure}[t]
\centering
\includegraphics[width=1\columnwidth]{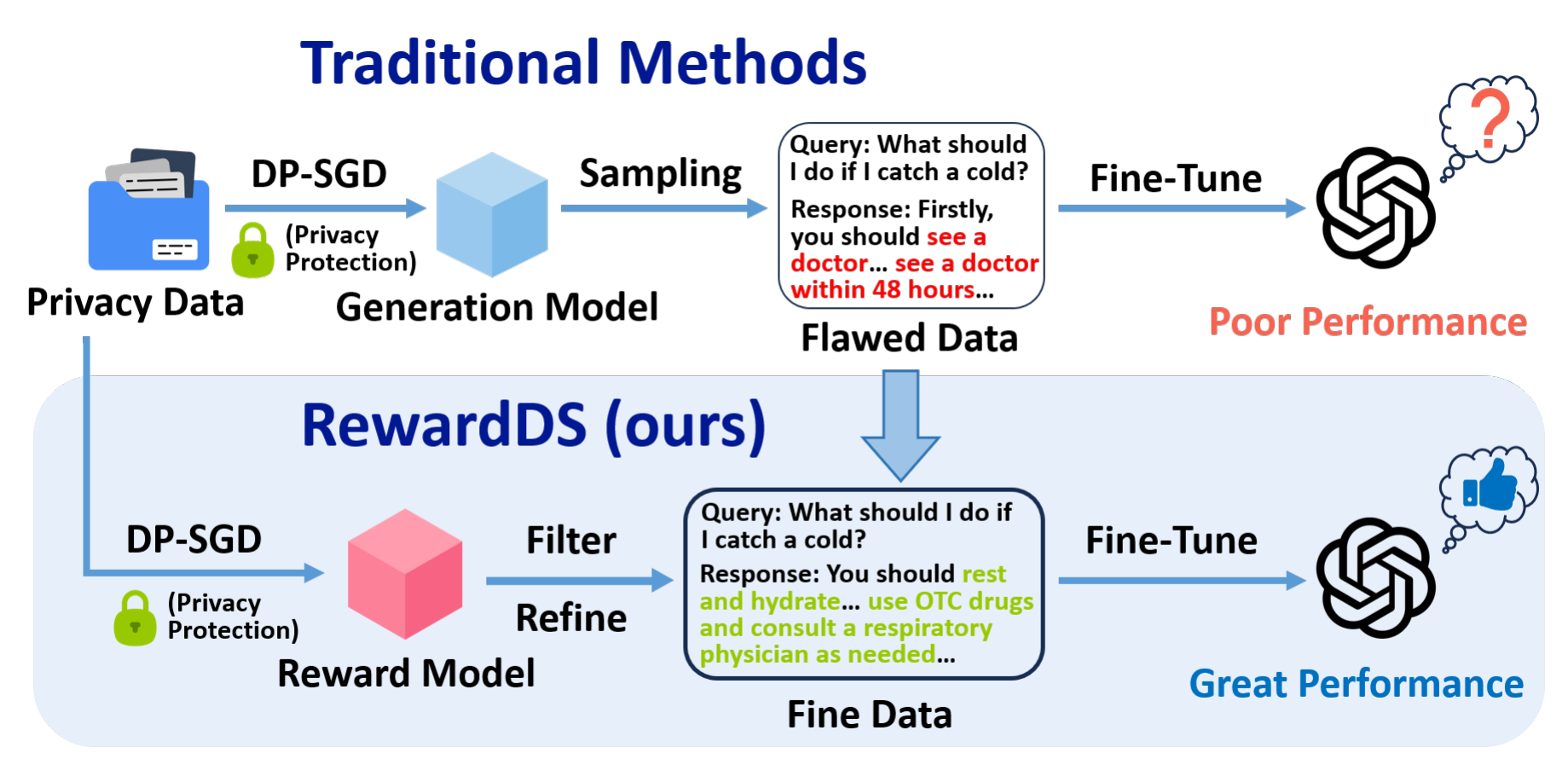}
\vspace{-15pt}
\caption{
Illustration of how \textit{RewardDS} overcomes the dilemma of traditional synthetic data methods.
The synthetic data directly sampled from the generation proxy model contain significant flaws, such as incoherent text or incomplete storylines, which are considered noise.
}
\vspace{-1.5em}
\label{fig:fig_intro}
\end{figure}

%Synthesizing data is a promising method to solve this challenge \citep{yue-etal-2023-synthetic, Kurakin2023HarnessingLM, yu2023training, mattern-etal-2022-differentially, Flemings2024DifferentiallyPK}
Prior works proposed data synthesis as a promising solution \citep{Kurakin2023HarnessingLM, yue-etal-2023-synthetic, yu2023training, mattern-etal-2022-differentially, Flemings2024DifferentiallyPK}. 
These approaches generate synthetic data to replace the private data and can be used for fine-tuning, thus ensuring privacy protection. 
Specifically, a generation proxy model is first trained on the private data, optimized by DP-SGD \citep{Abadi2016DeepLDP} to safeguard privacy. 
The generation proxy model then generates synthetic data for subsequent LLM training.
%Synthetic data is then sampled from this generation proxy model. 
% \revisezq{How to generate/sample synthetic data? Do you have prompt?}
However, due to the inherent randomness of the generation process, the synthetic data inevitably contain significant flawed one, including text incoherence or storyline incompleteness, which is considered as noise and leads to less effective LLM fine-tuning, as illustrated in Figure \ref{fig:fig_intro}.
% \revisezq{I think this figure does not show that the noise leads to decreased performance.}

% aims to ensure privacy in the original datasets by creating a new datasets utilizing Differential Privacy (DP) \citep{DBLP:journals/fttcs/DworkR14}. 
% The DP synthetic text methods are used widely in developing any downstream NLP system without adding extra privacy risks. 
% The state-of-the-art DP synthetic method is to \textit{finetune a client-side pretrained generative language models(LMs)} on private datasets \citep{yue-etal-2023-synthetic,Kurakin2023HarnessingLM} using DP-SGD \citep{Abadi2016DeepLDP}.
% The client-side LM generates synthetic responses for certain queries and transmits synthetic datasets to the server, then the server uses the synthetic data to fine-tune domain-specific LLMs. A series of methods have been proposed to prevent the direct use of private data by using synthetic data for substitution \citep{putta2022differentially,Lin2023DifferentiallyPS,Li2024SyntheticD}.

% 我们的contribution——多的reward model

To mitigate the noise, existing methods \citep{dayu2024privacy, Wang2022SelfInstructAL, Xie2024DifferentiallyPS} proposed to filter out flawed data by measuring its similarity to private data. 
\citet{Wang2022SelfInstructAL} use ROUGE-L similarity, while \citet{dayu2024privacy, Xie2024DifferentiallyPS} compute embedding similarity. 
However, these metrics fail to evaluate the synthetic data's effectiveness for domain-specific tasks. 
Alternative methods \citep{Wang2024KnowledgeSGPS, lifederatedSD24} attempt to distill the capabilities from the LLM on the server side into the generation proxy model to support domain-specific tasks.
However, since the LLM is not fine-tuned for these specific domains, the distillation provides limited benefit and does not effectively improve task performance.
We also present an illustrative example in Figure~\ref{fig:case_study}.

To more effectively mitigate noise in synthetic data, we propose \textit{RewardDS} (\textbf{Reward}-driven \textbf{D}ata \textbf{S}ynthesis), a novel privacy-preserving framework that improves synthetic data quality for the target LLM's privacy-preserving fine-tuning. 
RewardDS implements a two-stage quality control process, i.e., filtering and refinement, as illustrated in Figure \ref{fig:fig_intro}. 
%RewardDS trains an additional reward proxy model to guide the data synthesis process. 
%After the data synthesis process by the generation proxy model, \textit{RewardDS} leverages the reward model to filter and refine the synthetic data, as shown in Figure \ref{fig:fig_intro}. 
%Specifically, in addition to the generation proxy model, we fine-tune a reward proxy model on private data to assess data quality for domain-specific tasks, using DP-SGD to ensure user privacy. 
Specifically, we first train a reward proxy model on private data to assess data quality for domain-specific tasks, using DP-SGD to safeguard privacy. 
%After sampling the raw synthetic data from the generation proxy model, we apply the reward proxy model to judge the data and filter out those with low rewards, a process we call \textbf{Reward Guided Filtering}. 
Through \textbf{Reward Guided Filtering}, we apply the reward proxy model to assess synthetic data generated by the generation proxy model and remove samples with low reward scores.
%For synthetic data with low rewards, we introduce the \textbf{Self-Optimizing Refinement} module.  
%We generate multiple candidate responses for each synthetic query and compute their reward scores. 
% \revisezq{Do you only refine filtered out samples? Could you add a motivation of refining filtered out samples?}
Filtering alone may remove a large amount of data, leaving only a small fraction. 
Therefore, we aim to further refine the synthetic data to obtain more high-quality data.
Our \textbf{Self-Optimizing Refinement} module generates multiple candidate responses for each synthetic query and computes their rewards. 
The generation proxy model analyzes the highest and lowest scoring responses and then generates improvement feedback. 
%Then we provide both the highest and lowest rewarded responses to the reward proxy model to generate feedback on how to improve data quality. 
%We prompt the LLM \revisezq{generation proxy model? target LLM?} to refine the synthetic data based on this feedback. 
Based on this feedback, the target LLM refines the synthetic data following a refinement instruction.
The resulting high-quality, filtered, and refined synthetic data are then used to fine-tune the target LLM for domain-specific tasks. 
%Finally, we fine-tune the target LLM on the filtered and refined synthetic data for those domain-specific tasks. 

We conduct extensive experiments across various domain-specific generation tasks, including Medical Question Answering (QA), Legal QA, and Code Generation tasks. 
The results consistently demonstrate the effectiveness of our method in improving the quality of the synthetic data, achieving better performance while preserving privacy.
% 实验的结果
Our main contributions are summarized as follows:

\begin{itemize}[topsep=0.2em, partopsep=0em, itemsep=0.5em, parsep=0em, leftmargin=2em]
    \item We propose \textit{RewardDS}, a novel privacy-preserving fine-tuning framework that improves the quality of synthetic data by training a Reward Proxy Model on the client side to guide synthetic data generation. 
    \item We introduce the Reward Guided Filtering and Self-Optimizing Refinement modules to filter and refine the synthetic data, thereby enhancing its quality.
    \item We conducted extensive experiments across Medical QA, Legal QA, and Code Generation tasks to validate the effectiveness of our proposed framework.
\end{itemize}

%% file: related_work.tex
\section{Related Work}
In this section, we will introduce the related work on LLM privacy-preserving fine-tuning methods, which are currently divided into three categories: Anonymity-based methods, Encryption-based methods and Synthesis-based methods.
\vspace{-0.6em}

\paragraph{Anonymity-based methods.} Techniques like k-anonymity and adversarial anonymization can identify and anonymize private data. But they will significantly harm data quality for domain-specific LLM fine-tuning on the server side.
\citep{robin2024large, Sweeney1997GuaranteeingAW, Romanov2020NaturalTA}

\vspace{-0.6em}
\paragraph{Encryption-based methods.} Some approaches employ encryption techniques, such as Homomorphic Encryption (HE) or Secure Multi-Party Computation (SMPC), to protect private data. 
However, encrypting data and maintaining secure communication between server and client incur significant computational and time overhead, making these methods impractical in real-world scenarios.
\citep{frery2025private, lou2020glyph, you2025prifft}

\vspace{-0.6em}
\paragraph{Synthesis-based methods.}
Recent studies have explored using synthetic data with differential privacy (DP) guarantees as a substitute for private data in LLM fine-tuning. 
While this offers a practical and efficient solution, the synthetic data often contain noisy or flawed samples that significantly hinder LLM fine-tuning. 
Simple filtering based on text similarity is insufficient to effectively eliminate such noise.
\citep{Kurakin2023HarnessingLM,yue-etal-2023-synthetic,dayu2024privacy,hou2024pre,Wang2024KnowledgeSGPS,Wang2022SelfInstructAL}

Due to the limited space, a detailed introduction of the above works can be found in Appendix \ref{app:related_work}.

%% file: method.tex
\section{Problem Statement}
We consider a scenario where the client holds domain-specific data, such as patient's medical records, which contain sensitive information.  
Hence, directly transmitting those data to servers for LLM fine-tuning is not allowed. 
This private data typically is structured as \textit{Query}-\textit{Response} pairs, with both query and response containing confidential private information \citep{Wang2024KnowledgeSGPS}. 
The server, which hosts the target LLM, offers only API access while keeping model weights confidential, preventing clients from accessing or locally fine-tuning the model. 
While clients can fine-tune some lightweight LLMs within their computational constraints, these models have inherently weaker capabilities than the target LLM. 
This creates a critical challenge: how to leverage a client's private data to improve the server-hosted LLM's performance on domain-specific tasks while preserving privacy.
% , given that clients cannot locally fine-tune the target LLM due to inaccessibility of model weights. 

Existing synthesis-based methods utilize a lightweight \textbf{Generation Proxy Model} to generate safe synthetic data for fine-tuning the target LLM on the server \citep{yue-etal-2023-synthetic, dayu2024privacy}. 
However, the randomness of the generation process introduces significant noise in the synthetic data, potentially causing performance degradation. 
Therefore, our main goal is to \textbf{\textit{explore a more effective method for mitigating the noise in synthetic data, enabling better fine-tuning performance while maintaining user privacy}}.

\section{Method}
\begin{figure*}[!ht]
\centering
\includegraphics[width=0.95\textwidth]{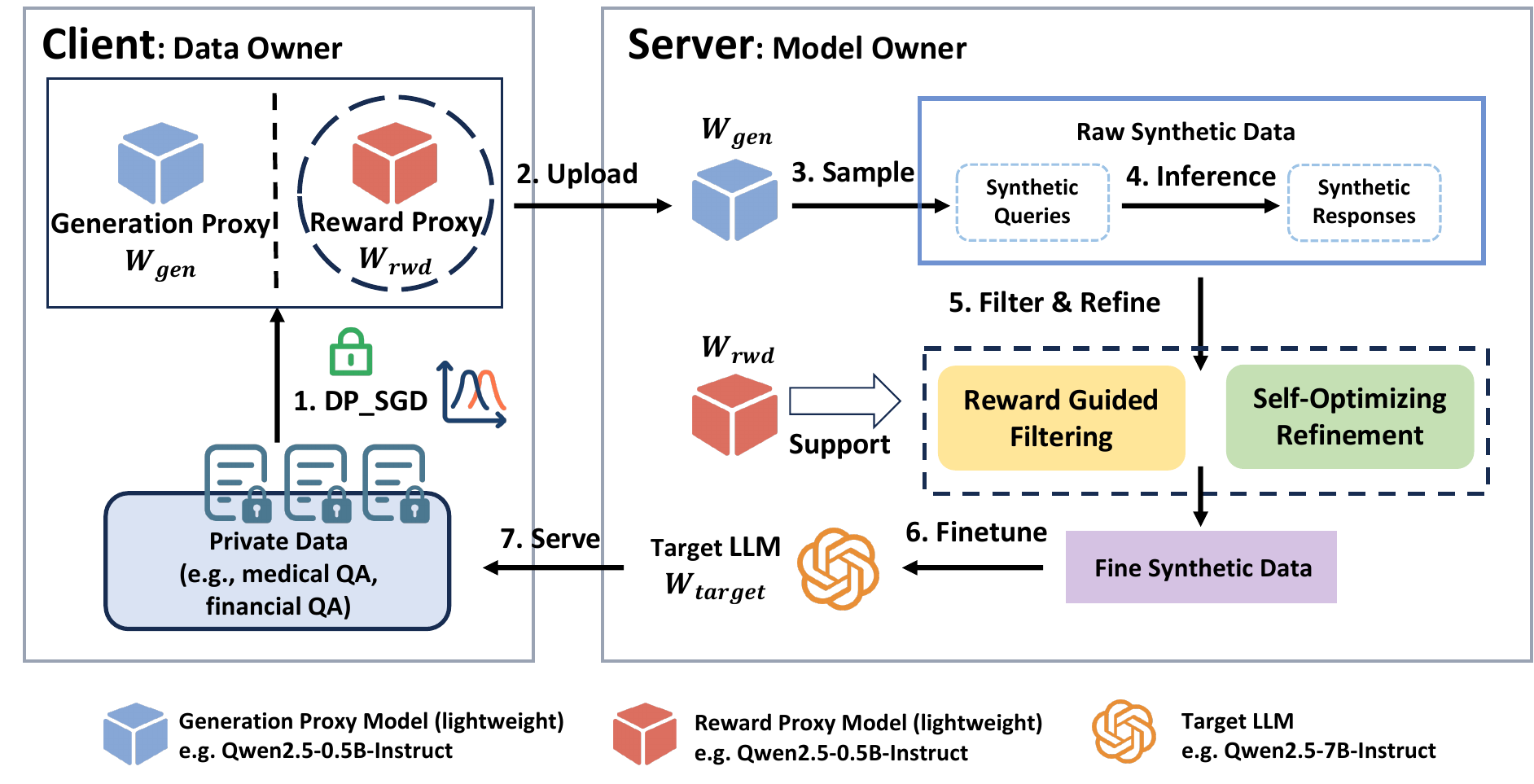}
% \vspace{-15pt}
\caption{
The overview of our \textit{RewardDS} framework.
The client uses DP-SGD to fine-tune two lightweight proxy models on privacy-sensitive data: the Generation Proxy Model $W_\text{gen}$ and the Reward Proxy Model $W_\text{rwd}$. 
Both proxy models are then sent to the server. 
The Generation Proxy Model is used to sample raw synthetic data, consisting of queries and responses. The Reward Proxy Model supports the \textbf{Reward Guided Filtering} and \textbf{Self-Optimizing Refinement} modules, which filter and refine the raw synthetic data to produce fine synthetic data. Finally, the target LLM $W_\text{target}$ is fine-tuned on the fine synthetic data and provides service to the client for domain-specific tasks.
}
\vspace{-0.7em}
\label{fig:algorithm}
\end{figure*}

To address the performance degradation caused by noise in synthetic data, we propose a novel framework, \textit{RewardDS} (\textbf{Reward}-driven \textbf{D}ata \textbf{S}ynthesis), as shown in Figure \ref{fig:algorithm}.
Our approach additionally trains a \textbf{Reward Proxy Model} on the client side. 
Then the reward proxy model filters and refines the synthetic data sampled from the generation proxy model through \textbf{Reward Guided Filtering} and \textbf{Self-Optimizing Refinement} modules on the server side. 
Both modules collaborate to enhance the quality of the synthetic data, driven by the reward signal from the reward model.
We will introduce the training process of the generation proxy model and reward proxy model in \cref{sec:Method_P1}. The details of reward guided filtering and self-optimizing refinement module are provided in \cref{sec:Method_P2}.

% \revisezq{Why need to fine-tune two proxy model?}

\setlength{\textfloatsep}{10pt}

\subsection{Client Side}
\label{sec:Method_P1}
\paragraph{Generation Proxy Model Training.}
% \revisezq{The motivation of fine-tuning a generation proxy model.}
The generation proxy model is responsible for generating safe synthetic data as a substitute for private data.
%The generation proxy model is directly trained on client's private data. 
%We utilize the generation proxy model to synthesize secure data which resembles the client's private data. 
Following \cite{yue-etal-2023-synthetic, dayu2024privacy, dayu2024dp, Kurakin2023HarnessingLM}, we fine-tune a generation proxy model on the client’s private data using the DP-SGD algorithm \cite{Abadi2016DeepLDP}. 
The backbone of generation proxy model should be lightweight due to limited computational resources on the client side, e.g., Qwen2.5-0.5B-Instruct \citep{yang2024qwen2}.  
The DP-SGD algorithm protects the privacy of the training data by injecting noise into the gradients during model training. 
This noise ensures that the inclusion or exclusion of any individual training sample has minimal impact on the fine-tuned model, thereby providing privacy protection. 
% \revisezq{The privacy budget of DP-SGD is defined as $(\epsilon, \delta)$, providing a theoretical measure of privacy protection.}
% \revisezq{This sentence is too detailed. Why need to mention it here. Can I put it into experiment?}

\paragraph{Reward Proxy Model Training.}
%Unlike the generation proxy model, which is fine-tuned directly on the client’s private data and used to synthesize similar but secure data, the reward model is responsible for evaluating the quality of the synthetic data. 
% \revisezq{Why you say fine-tune, I think it should be train. }
%We train the reward proxy model on delicatedly designed data. 
%Assuming the initial backbone model is $W_0$, the fine-tuned generation proxy model is $W_{gen}$, the fine-tuned reward proxy model is $W_{\text{rwd}}$ and the query-response pair from private data is ($Q, A_{\text{gold}}$).
%Following \citet{liu2024skywork}, we should construct a chosen response $A_c$ and a rejected response $A_r$ for each user query $Q$ before reward model training.
%We feed the user query $Q$ into $W_0$ and $W_{gen}$ to obtain corresponding responses $A_0$ and $A_{\text{gen}}$. 
%Then, we choose $A_{\text{gen}}$/$A_\text{gold}$ as the chosen response $A_c$ and $A_0$ as the rejected response $A_r$. 
%The reward proxy model is also lightweight as it is fine-tuned on the client side.
%We fine-tune it using DP-SGD to prevent privacy leakage. 
%We also adopt the lightweight model as the backbone model and fine-tune it using DP-SGD to prevent privacy leakage. 
% Qwen2.5-0.5B-Instruct
%We employ the Bradley-Terry model \citep{bradley1952rank} to define the training loss:
The reward model is responsible for evaluating the quality of the synthetic data. 
It should provide higher rewards for high-quality data while lower rewards for poor-quality data.  
Following standard reward model training practices \citep{liu2024skywork}, we train the reward proxy model using paired comparison data. 
Let $W_0$ denote the initial backbone model, $W_{gen}$ the fine-tuned generation proxy model, and $W_{\text{rwd}}$ the fine-tuned reward proxy model. 
For each query $Q$ from the private dataset with its gold response $A_{\text{gold}}$, we generate two responses: $A_0$ from $W_0$ and $A_{\text{gen}}$ from $W_{gen}$. 
We then create preference pairs by selecting either $A_{\text{gen}}$ or $A_{\text{gold}}$ as the chosen response $A_c$, with $A_0$ serving as the rejected response $A_r$.
The reward proxy model maintains a lightweight architecture for client-side deployment and is fine-tuned using differential privacy (DP-SGD) to prevent privacy leakage.

Following \citet{Ra2023DPO}, we define the training loss as:
\begin{equation}
\mathcal{L} = -\log \sigma\left( f_{\text{rwd}}(Q, A_c) - f_{\text{rwd}}(Q, A_r) \right),
\end{equation}
where $f_{\text{rwd}(\cdot)}$ represents the reward predicted by $W_\text{rwd}$.
This training loss encourages the reward model to assign higher scores to responses from the generation proxy model and gold responses compared to those from the initial backbone model. 

After training, both generation proxy model and reward proxy model are sent to the server. 
% We analyze the privacy preserving property of this transmission in \S \ref{sec:privacy_analysis}.

\subsection{Server Side}
\label{sec:Method_P2}

\paragraph{Synthetic Data Generation.}
% \revisezq{Since our goal is to fine-tune the LLM on the server, we should sample synthetic data from the generation proxy model. }
% \revisezq{Check Chinsese Comment.}
%ZZQ-Feb-14: 因果关系不明显。为什么用generation proxy model sample synthetic data 的动机讲清楚。
%First, following \citet{dayu2024privacy, Wang2024KnowledgeSGPS}, we prompt $W_\text{gen}$ to generate raw synthetic queries. 
%Second, we use $W_\text{gen}$ to generate synthetic responses for those queries. 
%We call these query and response pairs as raw synthetic data. 
Following \citet{dayu2024privacy, Wang2024KnowledgeSGPS}, we use $W_\text{gen}$ to generate both synthetic queries and their corresponding responses, collectively referred to as raw synthetic data. 
Although the generation proxy model $W_\text{gen}$ is trained on private data and learns domain-specific knowledge, the generation process of raw synthetic data is random and unstable. 
As a result, the raw synthetic data inevitably contains noisy samples, and fine-tuning the LLM directly on this data can lead to performance degradation.

\paragraph{Reward Guided Filtering.}
% A simple yet effective approach is to use the 
We leverage the reward proxy model $W_\text{rwd}$ to evaluate each synthetic data and filter out those with low rewards.
A lower reward indicates a higher likelihood of the synthetic data being noisy. 
We select only the top $\lfloor L/k \rfloor$ data, where $L$ is the total number of synthetic data and $k$ is the partition fold (Line \ref{line:filter1} in \cref{algo:main}).
To compensate for the reduced synthetic dataset size after filtering, we replicate the high-reward data to maintain the total data volume during the target LLM fine-tuning (Line \ref{line:filter2} in \cref{algo:main}). 

\paragraph{Self-Optimizing Refinement.}
%Relying solely on filtering does not fully utilize the entire synthetic dataset and may even lead to the LLM overfitting the small number of samples with high reward. 
%Given the self-reflective capabilities of LLMs \citep{aman2023self}, we explore refining the synthetic data during the LLM fine-tuning process. 
%Prior to LLM fine-tuning, we simultaneously generate $N$ candidate responses through the generation proxy model for each synthetic query (Line \ref{line:refine1} in \cref{algo:main}).
%And then the response with the highest reward, judged by the reward proxy model, is selected as the final synthetic response (Line \ref{line:refine2} in \cref{algo:main}). 

While filtering mitigates noise, it selects only a small subset of samples, potentially leading to overfitting on limited data.
Building on LLMs' self-reflection capabilities \citep{aman2023self}, we implement a dynamic data refinement strategy to improve low-reward samples, enhancing overall data quality.
Initially, for each synthetic query, we generate $N$ candidate responses rather than only one response using the generation proxy model (Line \ref{line:refine1} in \cref{algo:main}). 
The reward proxy model then selects the response with the highest reward score as the chosen response (Line \ref{line:refine2} in \cref{algo:main}). 
We directly fine-tune the target LLM $W_\text{target}$ on the chosen response (Line \ref{algo:line16} in \cref{algo:main}).
% Unlike greedy search generation, generating multiple responses allows for parallel exploration of different response trajectories.
% \revisezq{Here, you select a response with the highest reward as final synthetic response.}

After fine-tuning the target LLM $W_\text{target}$ for each epoch, we dynamically refine the synthetic data for the next epoch’s training. 
For each query’s $N$ candidate responses, we identify the lowest-reward responses and combine them with the highest-reward responses to form the rejected ($A_r$) and chosen ($A_c$) response pairs. 
The generation proxy model $M_{\text{gen}}$ analyzes these responses and provides feedback, highlighting the strengths of $A_c$ and weaknesses of $A_r$ (Line \ref{line:refine3}). 
This feedback, along with the original query, guides the target LLM $W_{\text{target}}$ to generate $N$ refined candidate responses (Line \ref{line:refine4}). 
Finally, the reward proxy model selects the highest-reward response from these refined candidates for the next epoch’s LLM fine-tuning (Line \ref{algo:lin20}).

% After fine-tuning the target LLM $W_\text{target}$ during each epoch, we refine the synthetic data with the help of the LLM itself for the next epoch training. 
% For each synthetic query's $N$ candidates, we also identify the lowest reward responses to combine with the highest reward responses as the rejected ($A_r$) and chosen ($A_c$) responses, respectively. 
% The generation proxy model $M_{\text{gen}}$ analyzes these responses and generate feedback which highlights the strengths of $A_c$ and weaknesses of $A_r$ (Line \ref{line:refine3}). 
% This feedback, combined with the original query, guides the target LLM $W_{\text{target}}$ to generate $N$ refined candidate responses (Line \ref{line:refine4}). 
% We use the reward proxy model to select the highest reward response from these refined candidate responses for the next epoch LLM fine-tuning.
% \revisezq{Here, you generate $N$ candidates. Is it contradictory?}

%For each synthetic query with $N$ candidate responses, we select the response with the lowest reward as the rejected response $A_r$ and the one with the highest reward as the chosen response $A_c$. 
%Both $A_c$ and $A_r$ are provided to the generation proxy model $M_{\text{gen}}$, which evaluates the strengths of the chosen response and the weaknesses of the rejected one, for generating feedback for further refinement (Line \ref{line:refine3}). 
%The user query, along with feedback, is then passed to the LLM $W_{\text{target}}$, which generates $N$ new candidate responses to achieve data refinement (Line \ref{line:refine4}).

The collaborative process between the reward-guided filtering and self-optimizing refinement modules is presented in \cref{algo:main}. 
The refinement instruction templates are provided in Appendix \ref{prompt_template_details}.
After the LLM is fine-tuned on the refined synthetic data, it can provide service to the client for those domain-specific tasks.

\begin{algorithm}[!t]
\SetAlgoLined
      \fontsize{8pt}{9pt}\selectfont
        \caption{\footnotesize \textit{RewardDS} based LLM Fine-tuning}
\label{algo:main}
    \KwIn{Synthetic query set $\mathcal{Q}_\text{query}$, number of synthetic query $L$, number of candidate responses $N$, partition fold $k$, generation proxy model $W_\text{gen}$, reward proxy model $W_\text{rwd}$, target LLM $W_\text{target}$, training epoch $T$ }
    \KwOut{The fine-tuned LLM $W_\text{target}^T$}

    \tcp{  Before Fine-tuning LLM}
    \For{each query $q \in \mathcal{Q}_\text{query}$}
    {   
        Generate candidate response set: $\{A_j\}_{j=1}^N \gets W_\text{gen}(q)$ \label{line:refine1}
    
        Predict the reward score: $\{s_j\}_{j=1}^N \gets W_\text{rwd}(q, A_j)$
        
        Select the best and the worst response: \newline
        \hfill
        $
        (A_c, A_r) \gets \left( A_{\argmax_j s_j},\ A_{\argmin_j s_j} \right)
        $ \label{line:refine2}
        
        Record the best reward score: $s_c \gets \max_j s_j$
    }
    
    \fontsize{8pt}{10pt}\selectfont
    Gather the initial synthetic dataset: $\mathcal{D}_0 \gets \{(q_i, A^i_c, A^i_r, s_c^i )\}_{i=1}^L$
     
    Sort $\mathcal{D}_0$ by reward: $\mathcal{D}_0^\text{sorted} \gets \{(q_i, A^i_c, A^i_r, s_c^i )\}_{i=1}^L$, where \newline
      \hfill $s_c^1 \geq \cdots \geq s_c^L$
      
    Partition $\mathcal{D}_0^\text{sorted}$ into $k$ folds: $\{\mathcal{D}_0^m\}_{m=1}^k \gets \text{split}(\mathcal{D}_0^\text{sorted}, k)$

    Extract top-$\lfloor L/k \rfloor$ samples: $\mathcal{D}_\text{high} \gets \mathcal{D}_0^1$ \label{line:filter1}
    
    Replicate subset to obtain the train set: $\mathcal{T}_0 \gets \bigoplus_{\lceil L/|\mathcal{D}_\text{high}| \rceil} \mathcal{D}_\text{high}$ \label{line:filter2}

    \fontsize{8pt}{9pt}\selectfont
    Determine score threshold $\tau$: 
    $
    \tau \gets \min_{s_c \in \mathcal{D}_\text{high}} s_c
    $

    \tcp{ During Fine-tuning LLM}
    Initialize target LLM: $W_\text{target}^0 \gets W_\text{target}$
    
    \For{iteration  $t=1$ \KwTo $T$}
    {  
        Fine-tune target LLM $W_\text{target}^{t-1}$ on $\{ (q, A_c) \in \mathcal{T}_{t-1}\}$ and get $W_\text{target}^t$ \label{algo:line16}

         \For{each query $q \in \mathcal{Q}_\text{query}$}
        {
        \fontsize{8pt}{10pt}\selectfont
            Generate feedback $\phi$: $\phi \gets W_\text{gen}(q, A_c, A_r)$ 
            \label{line:refine3}
            
            Re-generate the candidate response set: \newline
            $\{A_j\}_{j=1}^N \gets W_\text{target}^t(q, \phi)$
            \label{line:refine4}

            Predict the reward score, select the best and worst responses, and record the highest reward score to update $\mathcal{D}_{t-1}$, yielding $\mathcal{D}_t$. \label{algo:lin20}
        }

        Filter and get new training set $\mathcal{T}_t$: \newline
        $
        \mathcal{T}_t \gets \left\{(q, A_c, A_r, s_c) \in \mathcal{D}_t \mid s_c \geq \tau \right\}
        $
        
    }
    \KwRet{$M_\text{target}^t$}

\end{algorithm}

%% file: privacy.tex
\section{Privacy Analysis}
\label{sec:privacy_analysis}
The only transmitted contents between the client and server are the generation proxy model and the reward proxy model. 
Both models are fine-tuned on the private dataset using the DP-SGD algorithm \cite{Abadi2016DeepLDP}. 
According to the definition of differential privacy (DP) \citep{DBLP:journals/fttcs/DworkR14}, adversaries cannot infer any private data from the fine-tuned proxy models. 
Additionally, based on the post-processing property of the DP framework \citep{DBLP:journals/fttcs/DworkR14}, any further operations on the two proxy models will not cause privacy leakage. 
All subsequent operations on the server, including synthetic data generation, reward-guided filtering, and self-optimizing refinement, are privacy-preserving.
Moreover, we conduct Data Extraction Attack \citep{carlini2021extracting} and Membership Inference Attack \citep{yeom2018privacy, choquette2021label} on our method to empirically demonstrate its privacy protection capability in Section \ref{sec:em_privacy}.

We have fine-tuned two proxy models on the private dataset and the privacy budget of each fine-tuning is $(\epsilon, \delta$).
According to the sequential composition law of DP mechanism \citep{DBLP:journals/fttcs/DworkR14}, the total privacy budget of our framework is $(2\epsilon, 2\delta$).

%% file: experiments.tex
\section{Experiments}
\subsection{Experiments Setup}
\label{sec:setup}

\paragraph{Datasets.}
We evaluate our method across three domain-specific generation tasks using established datasets: Medical QA using HealthCareMagic-100k \citep{li2023chatdoctor}, Financial QA using fingpt-fiqa\_qa \citep{zhang2023instructfingpt}, and Code Generation using opc-sft-stage2 \citep{Huang2024OpenCoderTO}. 
%We evaluate our method on three domain-specific generation tasks: Medical Question-Answer (Medical QA), Financial Question-Answer (Financial QA), and Code Generation. 
%We utilize the HealthCareMagic-100k dataset \citep{li2023chatdoctor} for medical QA task, the fingpt-fiqa\_qa dataset \citep{zhang2023instructfingpt} for financial QA task and the opc-sft-stage2 dataset \citep{Huang2024OpenCoderTO} for code generation task. 
% \revisezq{Move all preprocess details into appendix. }
%ZZQ-Feb-11 免得reviewer质疑为什么要去掉low-quality samples，会质疑你为了自己方法好，改数据集。另外，appendix需要列出什么data才会判定为low-quality data，并给几个low-quality data的例子

\paragraph{Evaluation Metrics.} 
For the evaluation of the QA task, we employ the ROUGE-1 (R1), ROUGE-L (RL) \citep{lin-2004-rouge}, and Perplexity (PPL) \citep{yu2024ppl} as metrics. 
While automated metrics focus on lexical overlap and fluency, we also use LLM-Judge \citep{Lia2023judging} to provide a more comprehensive assessment of semantic accuracy and response quality. 
For the code generation task, we use Pass@1 and Pass@10 as evaluation metrics \citep{chen2021codex}.

\paragraph{Implementation Details.}
We use the Qwen2.5-0.5B-Instruct model \citep{yang2024qwen2} as the backbone for the generation/reward proxy model, and the Qwen2.5-7B-Instruct model as the target LLM on the server. 
During each DP-SGD fine-tuning process of both proxy models, we set the privacy budget to $(8, 1e^{-5})$. 
As a result, the total privacy budget for our method is $(16, 2e^{-5})$, according to the sequential composition law of the DP mechanism \citep{Abadi2016DeepLDP}.
For a fair comparison, we set the same privacy budget for all compared methods. 
The size of the synthetic dataset is always kept to twice that of the client's private data across all baselines. 
These settings align with established DP deployments such as Apple's QuickType and Google's models, as noted by \citet{Nils2023ana}.

More details on the datasets used and the implementation are provided in Appendix \ref{app:data_details} and Appendix \ref{app:imple_details}, respectively.

% \revisezq{Do you plan to fine-tune 13B on server side. Normally, reviewers want to see more backbones and more sizes.}
\subsection{Compared Methods.}
To demonstrate the effectiveness of our method, we consider several baselines for comparison:

\textbf{Vanilla LLM} refers to using a general-purpose LLM for domain-specific tasks without any domain adaptation or fine-tuning. 
\textbf{Locally Fine-tuning} refers to training a lightweight model locally on clients' private data. 
% \revisezq{why call it proxy model? I think the proxy should be deleted.}

\textbf{DP-Generation} \citep{Kurakin2023HarnessingLM} fine-tunes the generation proxy model on the client side using DP-SGD. This proxy model is then used to generate synthetic data, which are subsequently utilized to fine-tune target LLM on the server.
% \revisezq{carefully explain this method. In the explanation, there is no DP. But your method is DP-Generation. State clearly about the data generation and fine-tuning happens at the client or the server?}
\textbf{DP-Instruct} \citep{dayu2024privacy} additionally filters synthetic data based on text similarity before LLM fine-tuning;
\textbf{KnowledgeSG} \citep{Wang2024KnowledgeSGPS} distills the capacity from LLM into the generation proxy model to enhance its performance.

More details of the compared method are provided in Appendix \ref{app:compare}. 

\subsection{Main Results}

\begin{table*}
	\centering
        \def\arraystretch{1.16}
	\resizebox{0.96\textwidth}{!}{
    	\begin{tabular}{l 
       p{1cm}<{\centering} p{1cm}<{\centering} p{1cm}<{\centering}
       p{1cm}<{\centering} p{1cm}<{\centering} p{1cm}<{\centering}
       p{1.7cm}<{\centering} p{1.7cm}<{\centering} 
       % cc ccc cc
         } 
	\toprule
		  \multirow{2}{*}{Methods} 
          &  \multicolumn{3}{c}{Medical QA} &  \multicolumn{3}{c}{Financial QA} &  \multicolumn{2}{c}{Code Generation} \\ 
            \cmidrule(lr){2-9}
			& R1 $\uparrow$  & RL $\uparrow$ &  \multicolumn{1}{c|}{PPL $\downarrow$ }  
            & R1 $\uparrow$  & RL $\uparrow$ &  \multicolumn{1}{c|}{PPL $\downarrow$} &
            Pass@1 $\uparrow$ &
            Pass@10 $\uparrow$ \\
            
            % \cmidrule(lr){1-9}
            % \cmidrule(lr){1-1} \cmidrule(lr){2-4}  \cmidrule(lr){5-7}  \cmidrule(lr){8-9}
            \cmidrule(lr){1-9}

            \multicolumn{1}{l|}{Vanilla LLM} & 21.60 & 11.50 &  \multicolumn{1}{c|}{1.34} 
            & 23.91 & 11.72 & \multicolumn{1}{c|}{1.38}
            & 18.82 & 42.06\\
            
            \multicolumn{1}{l|}{Locally Fine-tuning} & \underline{23.82} & \underline{15.46} &  \multicolumn{1}{c|}{1.71} &
            13.26 & 10.19 & \multicolumn{1}{c|}{1.67} 
            & \underline{28.34} & 43.99 \\

            % \midrule
            % \cmidrule(lr){1-1} \cmidrule(lr){2-4}  \cmidrule(lr){5-7}  \cmidrule(lr){8-9}
            \cmidrule(lr){1-9}
            
            \multicolumn{1}{l|}{DP-Generation \citep{Kurakin2023HarnessingLM}} & 16.22 & 10.94 &  \multicolumn{1}{c|}{\underline{1.06}}
            & 14.97 & 11.20 & \multicolumn{1}{c|}{1.05}
            & 25.51 & 42.75 \\
            
            \multicolumn{1}{l|}{DP-Instruct \citep{dayu2024privacy}} & 11.94 & 8.44 & \multicolumn{1}{c|}{\textbf{1.04}} &
            14.06 & 10.76 & \multicolumn{1}{c|}{\underline{1.04}} 
            & 26.27 & 48.06 \\
            
            \multicolumn{1}{l|}{KnowledgeSG \citep{Wang2024KnowledgeSGPS}} & 20.28 & 10.74 & \multicolumn{1}{c|}{1.31} &
            \underline{24.14} & 12.33 & \multicolumn{1}{c|}{1.21} 
            & 23.93 & \underline{49.58} \\

            \rowcolor{lightgray!45}
            \multicolumn{1}{l|}{\textbf{RewardDS}} & \textbf{27.78} & \textbf{17.02} & \multicolumn{1}{c|}{1.17} &
            \textbf{24.42} & \textbf{14.96} & \multicolumn{1}{c|}{\textbf{1.02}} 
            & \textbf{32.41} & \textbf{49.99} \\
            
            \rowcolor{lightgray!20}
            \multicolumn{1}{l|}{w/o Reward Guided Filtering}  & 20.38 & 13.11 & \multicolumn{1}{c|}{1.28} & 
            17.93 & \underline{12.52} & \multicolumn{1}{c|}{1.25} 
            & 23.03 & 34.96 \\

            \rowcolor{lightgray!20}
            \multicolumn{1}{l|}{w/o Self-Optimizing Refinement} & 22.70 & 13.42 & \multicolumn{1}{c|}{1.36} &
            14.14 & 11.07 & \multicolumn{1}{c|}{1.18} 
            & 22.27 & 33.17 \\

            \bottomrule

		\end{tabular}
  
	}

 \caption{
Comparisons of our method with baselines across three domain-specific tasks: Medical QA, Financial QA, and Code Generation.
Higher values of ROUGE-1 (R1) and ROUGE-L (RL), and lower values of Perplexity (PPL) indicate better performance on the QA generation task. 
Higher values of Pass@1 and Pass@10 reflect better performance in the code generation task. 
Numbers in \textbf{bold} and \underline{underlined} represent the best and second-best results, respectively. 
% \revisezq{The results demonstrate that \textit{RewardDS} significantly outperforms the baselines.}
}
% \vspace{-0.7em}
\label{tbl:main}
\end{table*}

As shown in Table \ref{tbl:main}, \textit{RewardDS} outperforms all other baselines across the three domain-specific tasks, except for the PPL on the Medical QA task. 
DP-Instruct achieves marginally lower PPL in medical QA.
This may be attributed to the filtering strategy based on similarity, which could lead the target LLM to overfit on these highly similar samples.

The Vanilla LLM exhibits suboptimal performance across medical QA, financial QA, and code generation tasks, primarily due to the lack of domain-specific fine-tuning on private data. 
While Locally Fine-tuning a lightweight proxy model (with only 0.5B parameters) mitigates privacy concerns, the small model's limited capacity hinders its ability to effectively learn domain-specific knowledge, leading to subpar performance.

DP-Generation samples synthetic training data to fine-tune the target LLM on the server.  
However, due to the randomness inherent in the sampling process, the resulting synthetic data contains significant noise, which severely impairs the fine-tuning performance of the LLM on the server.
DP-Instruct attempts to filter the synthetic data by computing the similarity between the synthetic query and the private query. 
But, similarity alone cannot accurately reflect the quality of synthetic data, where higher similarity does not necessarily indicate better data quality.
KnowledgeSG distills the capabilities of the target LLM on the server into the generation proxy model for domain-specific tasks.
However, since the target LLM is not specifically fine-tuned for these tasks, the improvement through distillation is limited.
% the quality of the synthetic data is highly dependent on the target LLM’s capacity for the specific domain task. 
% If the target LLM performs poorly, the synthetic data will likely contain more noise, which could harm the performance of the proxy model. 
% KnowledgeSG performs relatively better on the financial QA task compared to the other two tasks, primarily because the Vanilla LLM performs well on financial QA, whereas it does not on the other tasks. 
% Only in the financial QA task, the performance of the Vanilla LLM surpasses that of the Locally Fine-tuned model, whereas this is not the case for the other tasks.

% We also present results after removing the Reward Guided Filtering and Self-Optimizing Refinement modules respectively. 
% We observe the decline in performance across all tasks when either module is removed, which highlights the effectiveness of these modules. 
% Without these modules, more noisy synthetic samples are included during LLM fine-tuning, resulting in performance degradation. 

We observe consistent performance declines across all tasks when either the Reward Guided Filtering or Self-Optimizing Refinement module is removed, highlighting the importance of both components.
Without these components, more noisy synthetic samples are used during LLM fine-tuning, leading to degraded performance.

\begin{figure}[!htbp]
  \centering
  % \vspace{-0.25em}

    \includegraphics[width=0.49\textwidth]{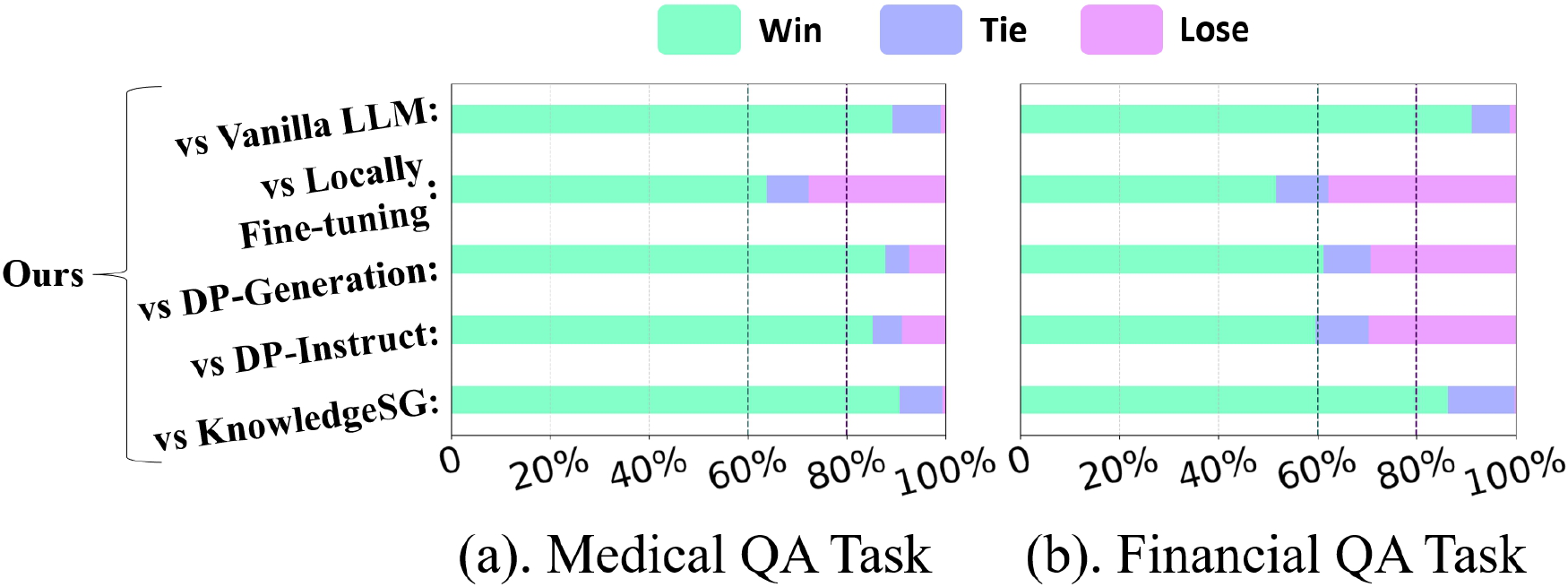}

  % \vspace{-0.25em}
  \caption{Using LLM-Judge to compare the outputs generated by our method with those of other baselines. 
  \textbf{Win} means our method outperformed the baselines, 
  \textbf{Tie} means the results were similar, 
  and \textbf{Lose} means our method performed worse than the baselines.
  }
  % \vspace{-0.75em}
  \label{fig:LLM-judge}
\end{figure}

In addition, following \citet{Lia2023judging}, we employ an LLM-Judger to assess the generated response from our method and those baseline approaches for QA tasks.
Specifically, we provide the LLM-Judger with responses from our method and those from baseline methods, prompting it to judge which response is better.
As shown in Figure~\ref{fig:LLM-judge}, our method consistently outperforms all baselines on both medical and financial QA tasks.
More implementation details are provided in Appendix~\ref{app:llm_judge}.
We also provide a case study in Figure~\ref{fig:case_study} and Appendix~\ref{app:case} to further demonstrate the effectiveness of our method.

\subsection{Empirical Privacy Protection Results}
\label{sec:em_privacy}
In this section, we implement Data Extraction Attack \citep{carlini2021extracting} and Membership Inference Attack \citep{yeom2018privacy, choquette2021label} on our method and baselines to empirically evaluate the privacy protection capability.
For those baselines, including DP-Generation/DP-Instruct/KnowledgeSG, the client only transfers the generation proxy model to the server.
In contrast, \textit{RewardDS} transfers both the generation proxy model and the reward proxy model.
Accordingly, the attack targets for the baselines are limited to the generation proxy model, while for \textit{RewardDS}, both models can be attacked.
We also provide the attack results of No Protection, which serves as the upper bound.

\begin{figure}[!htbp]
  \centering
  % \vspace{-0.25em}
% \raggedright
        \begin{minipage}{0.41\textwidth}
        \includegraphics[width=1\columnwidth]{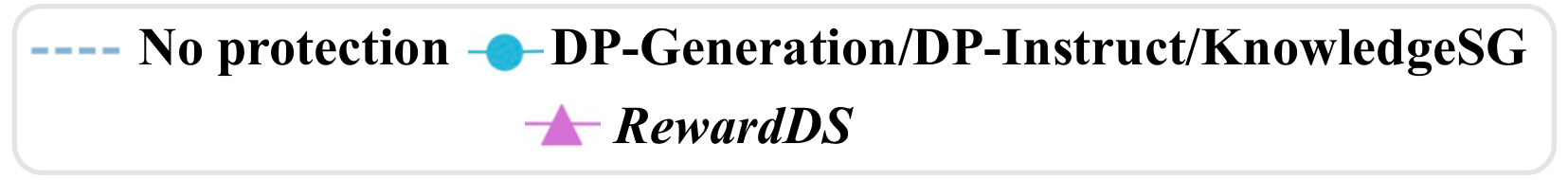}
        \end{minipage}
  \vspace{-1.3em}
  \vskip\baselineskip % 换行
  % 图片 1
  \raggedright
      \begin{minipage}{0.49\textwidth}
        \centering
        \includegraphics[width=1\columnwidth]{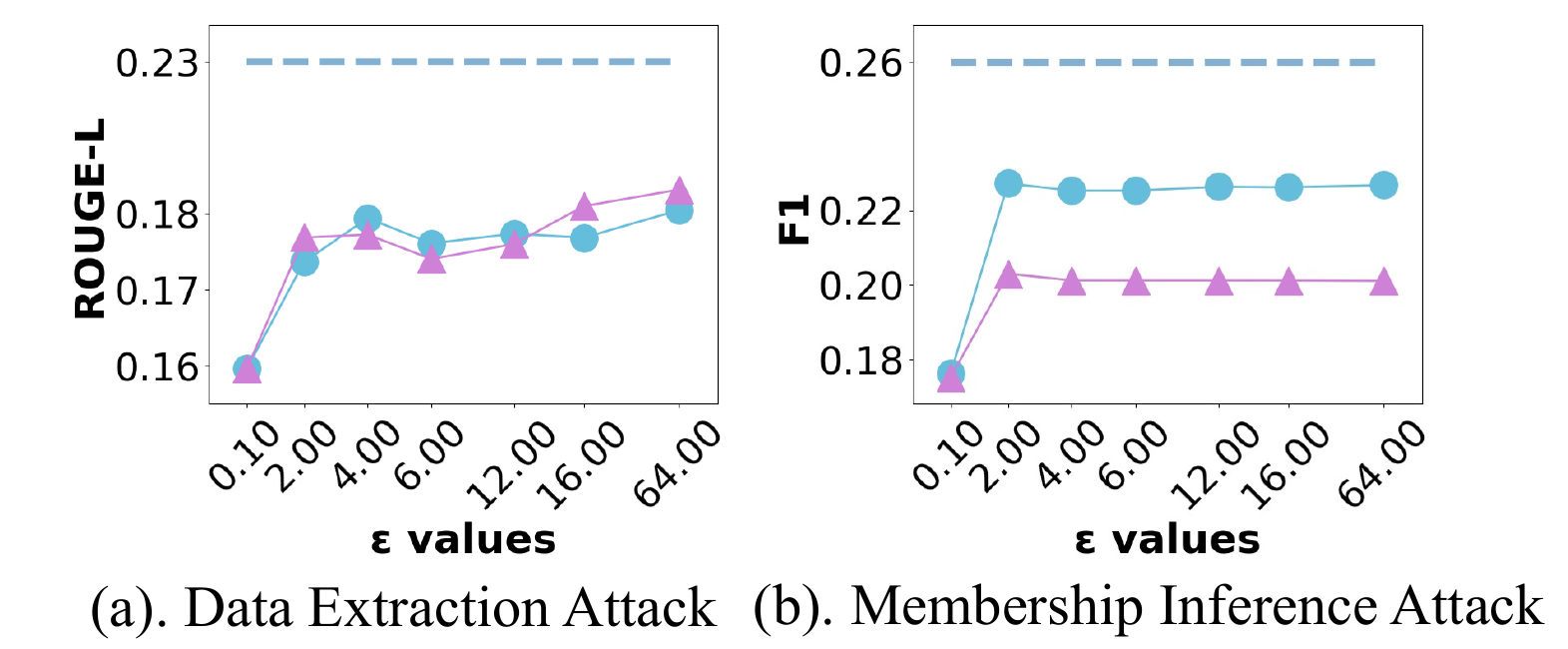}  
      \end{minipage}%
  % \vspace{-0.25em}
  \caption{Results of Data Extraction Attack and Membership Inference Attack for \textit{RewardDS} and all baselines under different privacy budgets $\epsilon$ on Medical QA task.}
  \label{fig:different_epislon}
\end{figure}

As shown in Figure~\ref{fig:different_epislon}, \textit{RewardDS} demonstrates superior privacy protection capacity compared to those baselines, as indicated by its comparable ROUGE-L scores and significantly lower F1 scores.
This is possibly due to \textit{RewardDS} allocating the privacy budget for both generation proxy model and reward proxy model, thereby reducing the privacy budget of each individual model and making them more difficult to be attacked.

Implementation details of these attack methods can be found in Appendix \ref{app:detail_privacy_protection_eval}.

\subsection{In-depth Analysis of \textit{RewardDS} Design}
% \revisezq{I prefer framing it as analysis rather than discussion}
Here, we provide more detailed analysis on the design and effectiveness of \textit{RewardDS}. 
% \begin{figure*}[!htbp]

%   \includegraphics[width=0.98\textwidth]{latex/fig/diss_2.pdf}
%     \label{fig:grow_legend}
%   \vspace{-0.7em}
%   \caption{
%   Effectiveness of \textit{RewardDS} design.
%   (a): Performance on medical QA with different privacy budget allocations for generation and reward proxy model training.
%   The allocation of `x + (16-x)' means the privacy budget for training the generation proxy model is set to x, while the reward proxy model is set to (16-x);
%   (b)/(c)/(d): Quality improvement of the synthetic data, on the Medical QA/Financial QA/Code Generation tasks, after applying the Self-Optimizing Refinement module multiple times.
%   }
%   \vspace{-1.0em}
%   \label{fig:different_epislon}
% \end{figure*}

\paragraph{Analysis 1:}\textit{Impact of RewardDS on Synthetic Data Quality and Downstream Performance.}

\begin{figure}[!htbp]

  \includegraphics[width=0.49\textwidth]{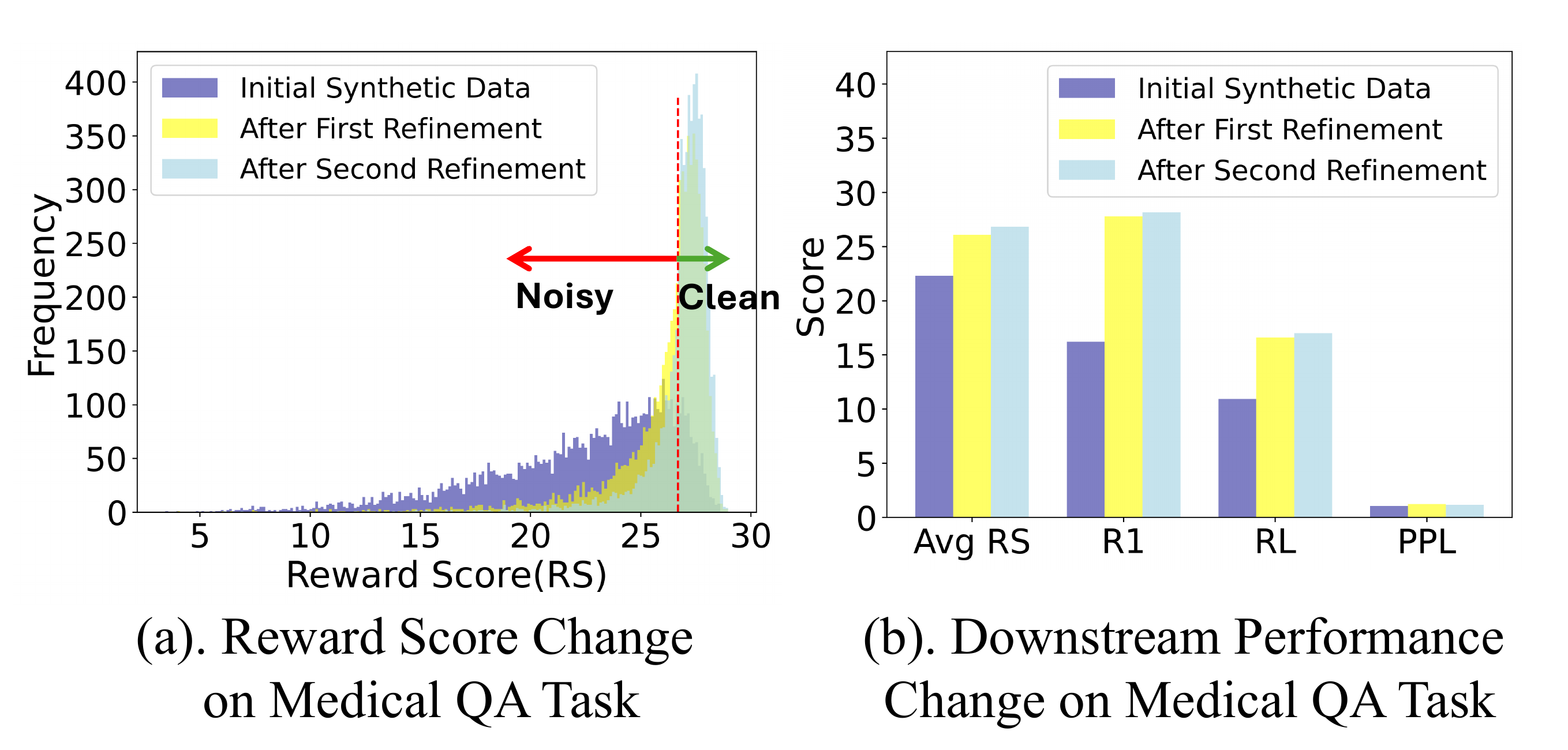}
     \vspace{-1.5em}
  \caption{
  Changes of Reward Score and Downstream Performance in \textit{RewardDS} for the Medical QA Task.
    \textbf{Avg RS} indicates the average reward score of the synthetic data, while \textbf{R1}, \textbf{RL}, and \textbf{PPL} represent the downstream performance metrics of the target LLM fine-tuned using these synthetic data.
  }
  \label{fig:change_visiable}
  \vspace{-0.9em}
\end{figure}

According to \cref{algo:main}, \textit{RewardDS} iteratively refines the synthetic data during each training epoch.
As shown in Figure\hyperref[fig:change_visiable]{~\ref{fig:change_visiable}(a)}, the reward score of synthetic data gradually increases with iterative refinement, indicating improved data quality.

We also evaluate the downstream performance of the target LLM on the Medical QA task after being fine-tuned on the synthetic data from different refinement stages.
The results in Figure\hyperref[fig:change_visiable]{~\ref{fig:change_visiable}(b)} show that the downstream performance of the fine-tuned LLM also improves significantly with the improvement of the synthetic data quality, strongly highlighting the effectiveness of \textit{RewardDS}.

\paragraph{Analysis 2:}\textit{Training Cost Analysis of RewardDS.}

As shown in Figure~\ref{fig:algorithm}, \textit{RewardDS} introduces additional modules, including Reward Proxy Model Training, Reward-Guided Filtering, and Self-Optimizing Refinement, into the privacy-preserving fine-tuning process of the server-side LLM.
We measure the additional time cost of these modules and compare it with that of the original modules: Generation Proxy Model Training and LLM fine-tuning.
As shown in Table~\ref{tbl:time_cost}, the additional time cost from our method accounts for \textbf{only 29.69\%} of the total time cost, with most of the time consumed by LLM fine-tuning modules.
This is primarily due to the use of a lightweight reward proxy model, making the associated modules highly efficient.
Overall, the additional time cost introduced by \textit{RewardDS} is minimal, strongly demonstrating the practicality of our method.

\begin{table}[!htbp]
    \centering
    \def\arraystretch{1} % 调整行间距
    % \resizebox{0.8\textwidth}{!}{ % 调整表格宽度为段落宽度的一半
    \resizebox{1\columnwidth}{!}{
    \scalebox{1}{
        \begin{tabular}{ l  |  cc}
        \toprule
         & \textbf{Time (min)} & \textbf{Percentage} \\
         \midrule
        \textbf{Initial Modules:} \cellcolor{lightgray!45} & 315 \cellcolor{lightgray!45} & 70.31\% \cellcolor{lightgray!45} \\
        Generation Proxy Model Training & 45 & 10.04\% \\
        \textbf{LLM fine-tuning} &  \textbf{270} & \textbf{60.26\%} \\
        \midrule
        \textbf{Additional Modules From \textit{RewardDS}:} \cellcolor{lightgray!45} & 133 \cellcolor{lightgray!45} & 29.69\% \cellcolor{lightgray!45} \\
        Reward Proxy Model Training & 49  & 10.94\% \\
        Reward Guided Filtering & 12 & 2.67\% \\
        Self-Optimizing Refinement & 72 & 16.07\% \\

        \bottomrule
        \end{tabular}
    }
    }
    % \vspace{-0.5em}
    \caption{
    Runtime of different modules in \textit{RewardDS} for privacy-preserving fine-tuning on Medical QA task. The most time-consuming module is marked in \textbf{bold}. 
    }
    \label{tbl:time_cost}
\end{table}

Moreover, we compare the overall training time and GPU memory usage of \textit{RewardDS} with other baselines in Table \ref{tbl:more_time_cost}.
As for training time cost, our method incurs only a small overhead, primarily from dynamically filtering and refining synthetic data to enhance the overall quality of these data.
For the GPU memory cost, we only introduce an additional lightweight reward proxy model with 0.5B parameters, which is relatively small and highly efficient.
Overall, with just a slight extra cost in training time and GPU memory, our method can achieves significant performance improvements, as shown in Table \ref{tbl:main}, making it highly practical for real world deployment.

\begin{table}[!htbp]
    \centering
    \def\arraystretch{1} % 调整行间距
    \resizebox{1\columnwidth}{!}{
    \scalebox{1}{
        \begin{tabular}[h]{l|cc}
            \toprule
            \textbf{Method} & \textbf{Time Cost (min)} & \textbf{Total Parameter (B)} \\
            \midrule
            \textbf{DP-Generation} & 315 & 7 + 0.5 \\
            \textbf{DP-Instruct} & 320 & 7 + 0.5 \\
            \textbf{KnowledgeSG} & 166 & 7 + 0.5 \\
            \midrule
            \textit{\textbf{RewardDS}} & 448 & 7 + 0.5 + 0.5 \\
            
            \bottomrule
        \end{tabular}
    }
    }
    \caption{Comparison of the time cost and the GPU memory cost between our method and the other baseline methods. 
    The GPU memory cost are measured by the total model parameter amount.}
    \label{tbl:more_time_cost}
\end{table}

\paragraph{Analysis 3:}\textit{\textit{RewardDS} vs Filtering according to Reward Score.}

According to Figure\hyperref[fig:change_visiable]{~\ref{fig:change_visiable}(a)}, the initial synthetic data contains substantial samples with low reward scores. 
One straightforward strategy is to filter out these low-quality samples to reduce noise. 
As shown in Table~\ref{tbl:com_filer}, simply selecting Top 50\% synthetic data for fine-tuning can improve the overall data quality and slightly enhance downstream performance.
However, filtering also reduces the size of the training set. 
As more data is discarded, the downstream performance begins to drop due to the limited training data. 

In contrast, \textit{RewardDS} applies Self-Optimizing Refinement to improve the quality of low reward samples instead of discarding them.
\textbf{It can significantly enhance data quality while maintaining a stable dataset size}. 
Consequently, \textit{RewardDS} achieves superior downstream performance, as demonstrated in Table~\ref{tbl:com_filer}.

%We also compute the reward score for all newly generated synthetic data and visualize the quality change after multiple refinements. 
% A higher reward score indicates better synthetic data quality. 
%As shown in subfigures (b), (c), and (d) of Figure \ref{fig:different_epislon}, \textbf{the synthetic data quality improves gradually with multiple applications of the self-optimizing refinement module}. 
% Figures \ref{fig:different_epislon}(b-d) demonstrate that \textbf{synthetic data quality improves gradually through iterative refinement}, explaining our method's superior performance.

\begin{table}[!htbp]
    \centering
    \def\arraystretch{1} % 调整行间距
    % \resizebox{0.8\textwidth}{!}{ % 调整表格宽度为段落宽度的一半
    \resizebox{1\columnwidth}{!}{
    \scalebox{1}{
        \begin{tabular}{ l  |  ccc}
        \toprule
         & \textbf{Data Count $\uparrow$} & \textbf{Avg RS $\uparrow$} & \textbf{Downstream RL $\uparrow$}  \\
         \midrule
        \textbf{Raw Synthetic Data} & \textbf{6420} & 22.30 & 10.94 \\
        -- Select Top 80\% & \underline{5136} & 24.05 & 11.09 \\
        -- Select Top 50\% & 3210 & 25.66 &  \underline{12.43} \\
        -- Select Top 30\% & 1926 & 26.53 & 11.19 \\
        -- Select Top 10\% & 642 & \textbf{27.43} & 6.82 \\
        \midrule
        \textbf{Refinement by \textit{RewardDS}} & \textbf{6420} \cellcolor{lightgray!45} & \underline{26.82} \cellcolor{lightgray!45} &  \textbf{17.02} \cellcolor{lightgray!45}  \\
        
        \bottomrule
        \end{tabular}
    }
    }
    % \vspace{-0.5em}
    \caption{
    Comparison between filtering synthetic data only based on Reward Score (RS) and iterative refinement using \textit{RewardDS} on Medical QA task.
    Higher \textbf{Avg RS} indicates better overall data quality.
    The best results are shown in \textbf{bold}, and the second-best results are \underline{underlined}.
    }
    \label{tbl:com_filer}
    \vspace{-0.9em}
\end{table}

\paragraph{Other Analysis:}
Furthermore, we analyze \textbf{the impact of different privacy budget allocations} for our method in Appendix~\ref{app:more_analysis}.
Those results in Figure \ref{fig:privacyA} also clearly demonstrate the effectiveness and robustness of our method.

\subsection{Hyperparameter Analysis}
\label{sec:hyper}
\begin{figure}[!htbp]
  \centering
  % \vspace{-0.25em}
% \raggedleft
        \begin{minipage}{0.40\textwidth}
        \includegraphics[width=1\columnwidth]{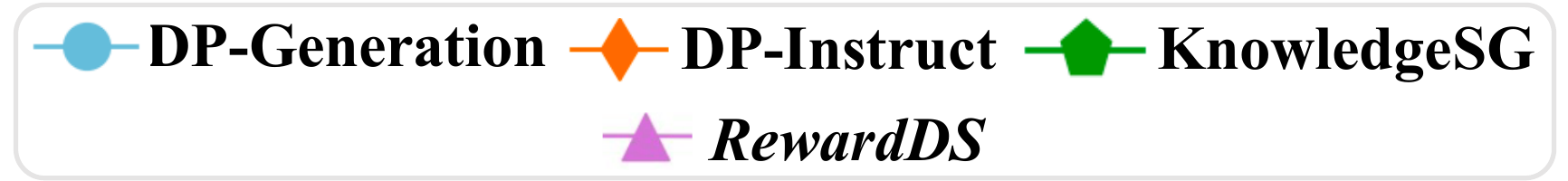}
        \end{minipage}
  \vspace{-1em}
  \vskip\baselineskip % 换行
  % 图片 1
  \raggedright
      \begin{minipage}{0.49\textwidth}
        \centering
        \includegraphics[width=1\columnwidth]{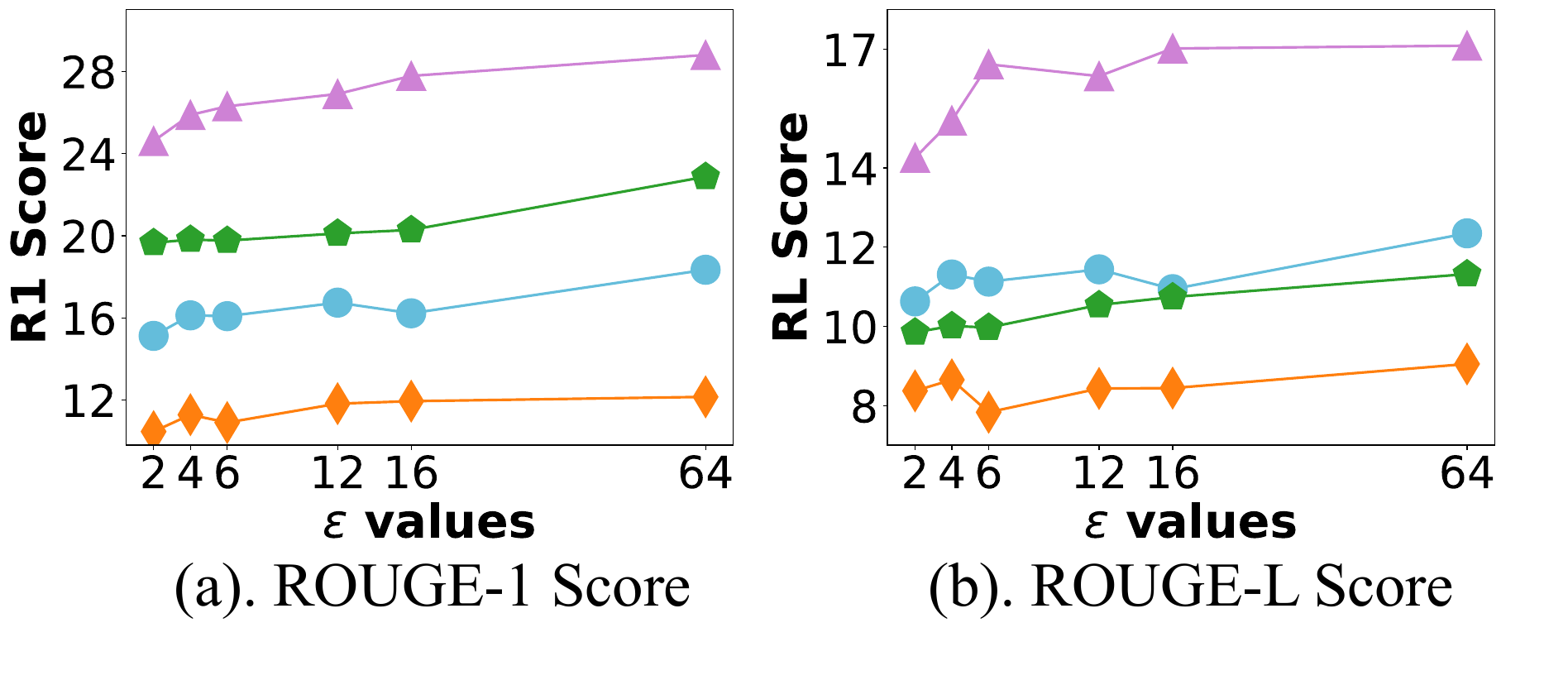}  
      \end{minipage}%
  \vspace{-0.9em}
  \caption{Performance of \textit{RewardDS} and baselines on the Medical QA task under different privacy budgets $\epsilon$ for the Medical QA task.
  The performance is evaluated using ROUGE-1 (\textbf{R1}) and ROUGE-L (\textbf{RL}) scores.}
  \label{fig:hyper_epislon}
\end{figure}

In this section, we conduct hyperparameter analysis to further prove the effectiveness of our method.
\paragraph{Privacy Budget $\epsilon$.}
We evaluate the performance of our method and baseline approaches under varying privacy budgets $\epsilon$.
As shown in Figure~\ref{fig:hyper_epislon}, our method consistently outperforms all baselines, not only under the commonly used setting of $\epsilon = 16$, but also under stricter privacy budgets (e.g., $\epsilon = 2$, $4$, and $6$).
These results demonstrate the superior performance and broad applicability of our method.

\paragraph{Synthetic Data Count $L$.}
To further validate the effectiveness of our method, we compare its performance with baseline methods under different synthetic data count $L$.
As shown in Table \ref{tbl:quantity}, RewardDS consistently outperforms all baselines across different data counts, demonstrating its robustness and effectiveness.
Notably, even when the synthetic data—for example is extremely scarce, only 1\% (642 samples), our method still maintains competitive performance while other baselines suffer from performance degradation.
This superiority is mainly attributed to our self-optimizing refinement module, which effectively improves the quality of those limited synthetic data and can maximize their utility in contrast to the baselines.

\begin{table*}[!htbp]
    \centering
    \def\arraystretch{1.2} % Adjust row spacing
    \resizebox{0.95\textwidth}{!}{ % Control table width
        \large
        \begin{tabular}[h]{l|ccccc}
            \toprule
            \textbf{Synthetic Data Count} & \textbf{642(1\%)} & \textbf{1926(30\%)} & \textbf{3210(50\%)} & \textbf{5136(80\%)} & \textbf{6420(100\%)} \\
            \cmidrule(lr){1-6}
            \textbf{DP-Generation} \citep{Kurakin2023HarnessingLM} & 6.71 & 9.55 & 10.60 & 10.30 & 10.94 \\
            \textbf{DP-Instruct} \citep{dayu2024privacy} & 6.63 & 8.00 & 7.99 & 8.45 & 8.44 \\
            \textbf{KnowledgeSG} \citep{Wang2024KnowledgeSGPS} & 9.43 & 10.21 & 10.54 & 10.92 & 10.74 \\ 
            
            \midrule
            \textbf{RewardDS} & \textbf{16.46} & \textbf{17.17} & \textbf{17.39} & \textbf{17.34} & \textbf{17.02} \\
            \bottomrule
        \end{tabular}
    }
    % \vspace{-0.5em}
    \caption{Performance comparison between \textit{RewardDS} and the other baseline methods under different synthetic data count for the Medical QA task.}
    \label{tbl:quantity}
\end{table*}

\paragraph{More Hyperparameters.}
We also analyze more hyperparameters in our method described in \cref{algo:main}, including the number of folds $k$ and the number of candidate responses $N$.
As shown in Figure \ref{fig:medical_hyper}, \ref{fig:fin_hyper} and \ref{fig:code_hyper}, our method remains effective and robust across different hyperparameter settings on three domain-specific tasks.
More detailed analyses can be found in Appendix \ref{app:hyper}.

\subsection{Generalization to More Proxy Models and LLMs}

We also evaluate the performance of our method when using different combinations of generation proxy models and reward proxy models, as shown in Table \ref{tbl:heterogeneity}.
To explore the impact of proxy model heterogeneity, we evaluate the combination of Qwen-0.5B + LLaMA3-1B.
To explore the impact of proxy model scale, we consider the combination of Qwen-0.5B + Qwen-1.5B.
As shown in the table, using Qwen-0.5B as the generation proxy model and Qwen-1.5B as the reward proxy model achieves the best performance. 
This is mainly because both proxy models share a similar architecture, which ensures more effective collaboration between them. 
Moreover, a larger reward proxy model can more effectively enhance the quality of synthetic data and achieve better performance, which underscores the critical role of the reward proxy model.

\begin{table}[!htbp]
    \centering
    \def\arraystretch{1} % Adjust row spacing
    \resizebox{1\columnwidth}{!}{ % Control table width
        \begin{tabular}[h]{l l|ccc}
            \toprule
            \textbf{Generation Proxy} & \textbf{Reward Proxy} & \textbf{RL Score} & \textbf{R1 Score} & \textbf{PPL} \\
            \cmidrule(lr){1-5} 
            Qwen-0.5B & Qwen-0.5B & 17.02 & 27.78 & 1.17 \\
            Qwen-0.5B & Llama3-1B & 17.69 & 28.71 & 1.21 \\
            Qwen-0.5B & Qwen-1.5B & \textbf{18.11} & \textbf{30.30} & 1.18 \\
            Llama3-1B & Qwen-0.5B & 12.52 & 19.35 & \textbf{1.15} \\
            Qwen-1.5B & Qwen-0.5B & 17.44 & 29.64 & 1.24 \\
            \bottomrule
        \end{tabular}
    }
    % \vspace{-0.5em}
    \caption{Performance of our method under different combinations of generation proxy models and reward proxy models for the medical QA task.}
    \label{tbl:heterogeneity}
\end{table}

Moreover, we evaluate the performance of our method with different server-side LLMs, such as Llama-2-7B-chat-hf \cite{metaai2023llama2} and Qwen2.5-14B-Instruct \cite{yang2024qwen2} in Appendix~\ref{app:more_analysis}.
As shown in Table \ref{tbl:more_back}, our method consistently outperforms other baselines across various LLMs, demonstrating its effectiveness and robustness.

%% file: conclusion.tex
\section{Conclusion}
We propose a novel privacy-preserving framework, \textit{RewardDS}, to mitigate noise in synthetic data during LLM privacy-preserving fine-tuning. 
Specifically, \textit{RewardDS} fine-tunes a reward model and leverages the reward signal to guide the synthetic data generation process. 
During the data synthesis process, \textit{RewardDS} employs the collaboration of Reward Guided Filtering and Self-Optimizing Refinement modules to filter and refine synthetic data, mitigating noise. 
We conduct extensive experiments across medical QA, legal QA, and code generation tasks. 
The results consistently demonstrate the effectiveness of \textit{RewardDS} for privacy-preserving LLM fine-tuning.

\newpage
\section*{Limitations}
% While \textit{RewardDS} has demonstrated its effectiveness in medical QA, legal QA, and code generation tasks, it incurs additional training costs for the reward proxy model. 
% Although the model is lightweight, it still requires extra computational resources.

Due to computational resource constraints, we applied LoRA fine-tuning on the Qwen2.5-14B-Instruct model to validate our method, as discussed in Appendix \ref{sec:ext_LLM}. 
Full-parameter fine-tuning may yield even better performance. 

Future work will investigate larger LLM backbones to further validate the effectiveness of our method on models with greater parameter scales.

Moreover, although our method is already time-efficient, we plan to further improve efficiency by exploring lightweight training approaches, such as Low-Rank Adaptation (LoRA) and prefix tuning, during the fine-tuning of both the generation proxy model and the reward proxy model.

\section{Ethics Statement}
We adhere to the ACL Ethics Policy and all of our research is based on publicly available repositories and datasets. 
In the \textit{RewardDS} framework, we uphold strict ethical standards to protect user privacy and ensure data security. 
The datasets used, covering medical QA, financial QA, and code generation domains, are publicly available and free of personally identifiable information, minimizing privacy risks.
Our methodology does not access or reconstruct identifiable data, safeguarding individual privacy rights.

However, as our study involves multiple LLMs, such as Llama and Qwen, the findings may be influenced by the inherent biases, linguistic patterns, and assertiveness of these models.

%% file: appendix.tex
\clearpage
\newpage
\appendix

\section*{Appendix Overview}
The appendix is organized into two parts: Appendix~A--D provide related work and the main experimental setup of \textit{RewardDS}, while Appendix~E--J present additional experimental results to further demonstrate the effectiveness of \textit{RewardDS}.

\section{Detailed Related Works}
\label{app:related_work}
In this section, we provide a more detailed introduction to the related works on LLM privacy-preserving fine-tuning methods, including Anonymity-based methods, Encryption-based methods and Synthesis-based methods.

\subsection{Anonymity-based methods}
Anonymity-based methods aim to identify and remove user-specific private information from private data to enable privacy-preserving LLM fine-tuning.
\citet{Sweeney1997GuaranteeingAW} achieves k-anonymity by dynamically identifying user-specific private information and applying substitution or removal to protect it.
\citet{Romanov2020NaturalTA} employs a transformer framework with attention mechanisms to enhance anonymization performance.
\citet{robin2024large} proposes an adversarial anonymization approach that leverages one LLM to anonymize user privacy while using another LLM to detect privacy information, iteratively improving the anonymization performance.

All of the above anonymity-based methods require detecting and removing user privacy, which will make the data incoherent and incomplete, thereby reducing its quality for downstream LLM fine-tuning.

\subsection{Encryption-based methods}

Encryption-based methods focus on applying encryption to the private data and maintaining secure communication between client and server to transmit the private data.
\citet{lou2020glyph} applies fully Homomorphic encryption to protect private data, enabling privacy security while maintaining comparable model performance after fine-tuning on the encrypted data.
\citet{frery2025private} combines the Low-Rank Adaptation technique and homomorphic encryption to improve the efficiency of LLM privacy-preserving fine-tuning.
\citet{you2025prifft} introduces the hybrid secret sharing algorithm by combining arithmetic secret sharing (ASS) and function secret sharing (FSS) to achieve secure computation during LLM privacy-preserving fine-tuning.

However, current encryption-based methods still require numerous time and resources for encrypting private data and ensuring secure communication, making them impractical for real-world applications.

% \citep{Kurakin2023HarnessingLM,yue-etal-2023-synthetic,dayu2024privacy,hou2024pre,Wang2024KnowledgeSGPS,Wang2022SelfInstructAL}
\subsection{Synthesis-based methods}
Synthesis-based methods have recently emerged as a more practical and reliable approach, which leverages synthetic data as a substitute for private data in LLM fine-tuning to balance utility and privacy.

\citet{Kurakin2023HarnessingLM,yue-etal-2023-synthetic} propose using DP-SGD to locally fine-tune a lightweight model on the client as a generation proxy. 
This proxy model is used to generate synthetic data without privacy risks.
The server utilizes the synthetic data to fine-tune the LLM, achieving privacy-preserving training.
Considering that those synthetic data often contain numerous incoherent and flawed samples, \citet{dayu2024privacy, hou2024pre} filter out low-quality data by measuring similarity between synthetic and private data. 
Alternatively, \citet{Wang2024KnowledgeSGPS, Wang2022SelfInstructAL} avoid sampling synthetic data from the generation proxy model, instead using LLM on the server to improve the proxy model by distillation.

Nevertheless, only text similarity is too surface-level to accurately assess the quality of synthetic data for domain-specific tasks. 
Moreover, since the server-side LLM is not fine-tuned for domain-specific tasks, its ability to enhance the generation proxy model through distillation is limited.

\section{Details of Datasets}
\label{app:data_details}

\begin{table*}[!b]
    \centering
    \def\arraystretch{1.7} % Adjust row spacing
    \resizebox{0.85\textwidth}{!}{ % Control table width
        \large
        \begin{tabular}[h]{l l|cccc}
            \toprule 
           \textbf{Task} & \textbf{Dataset} & \textbf{Private Train Set} & \textbf{Dev Set} & \textbf{Test Set} & \textbf{Sampling Synthetic Data} \\
            \cmidrule(lr){1-6}
            \textbf{Medical QA} & HealthCareMagic-100k & 3210 & 112 & 1683 & 6420 \\
            \textbf{Financial QA} & fingpt-fiqa\_qa & 1693 & 18 & 1711 & 3386 \\
            \textbf{Code Generation} & opc-sft-stage2 & 1497 & 79 & 1449 & 2994 \\
            \bottomrule  
        \end{tabular}
    }
    \vspace{-0.5em}
    \caption{The dataset statistics of the medical QA, financial QA and code generation task. All train set is hold by the client and is regard as the private data. 
    The size of sampling synthetic data is two times of the size of the private train set.}
    \label{tbl:statistics}
\end{table*}

To evaluate the performance of the compared methods on domain-specific tasks, we focus on three tasks: Medical Question-Answering (QA), Financial QA, and Code Generation. 
For the medical QA task, we use the HealthCareMagic-100k dataset \citep{li2023chatdoctor}; for the financial QA task, we use the fingpt-fiqa\_qa dataset \citep{zhang2023instructfingpt}; and for the code generation task, we use the opc-sft-stage2 dataset \citep{Huang2024OpenCoderTO}.

As \citet{dong2024gene} points out, these public datasets suffer from a "data contamination" issue, where some of the data may have been used to train LLMs on the server, causing the models to memorize it and leading to unnaturally high performance.
Moreover, the initial datasets are highly redundant, containing many similar samples. 
To accurately assess the domain-specific performance of different baselines, we should pre-process these datasets. 
To be specific, firstly, we evaluate the dataset using the Qwen2.5-7B-Instruct model \cite{yang2024qwen2} and exclude samples with high accuracy, as higher accuracy suggests these samples may have been part of the LLM's training data and are thus contaminated.

After addressing the contamination issue, we use the Sentence-T5-Base model \cite{ni2022sen} to compute embeddings for each sample and calculate their similarity. 
This allows us to remove highly similar samples, ensuring deduplication. 
The pre-processed dataset is then split into private train set, dev set, and test set, with the detailed statistics shown in Table \ref{tbl:statistics}. For fair comparison across all methods, we control the size of our sampled synthetic dataset to be twice the size of the private training set, as shown in Table \ref{tbl:statistics}.

\section{Compared Methods}
\label{app:compare}
Here, we will provide more detailed introductions to all compared methods:

% \textbf{Vanilla LLM} involves users directly utilizing the LLM on the server for those domain-specific tasks;
% \textbf{Locally Fine-tuning} involves users fine-tuning a lightweight proxy model locally for those tasks. 

\paragraph{Vanilla LLM:} Vanilla LLM directly uses the LLM (Qwen2.5-7B-Instruct) on the server for those domain-specific tasks.
% 直接利用服务器上的LLM来完成那些特定于领域的任务
% Qwen 7b

\paragraph{Locally Fine-tuning:} Locally Fine-tuning fine-tunes a lightweight model (Qwen2.5-0.5B-Instruct) using the private data on the client for those domain-specific tasks. 
% 用户在本地为这些任务微调轻量级代理模型
% Qwen 0.5b

\paragraph{DP-Generation:} As proposed by \citet{Kurakin2023HarnessingLM}, DP-Generation first uses DP to fine-tune a lightweight model (Qwen2.5-0.5B-Instruct) as the Generation Proxy Model on the client side.
Then, it transmits the Generation Proxy Model to the server for synthetic data sampling. 
% For each task, create a prefix $p =$
% \texttt{"[TaskName] [LabelName\textsubscript{y}]"} as the model input and autoregressively sampled at least the same amount of synthetic data as in the original training dataset. 
Then, the synthetic data is used to fine-tune the LLM (Qwen2.5-7B-Instruct) on the server for those domain-specific tasks.
% \textbf{DP-Generation} \citep{Kurakin2023HarnessingLM} fine-tunes the generation proxy model and generates synthetic data for LLM fine-tuning; 
% 对生成代理模型进行微调，并生成用于LLM微调的合成数据
% 微调Qwen 0.5b模型，再微调server 7B

\paragraph{DP-Instruct:} Compared to DP-Generation, DP-Instruct \citep{dayu2024privacy} introduces a filtering step to improve the quality of synthetic data.
After sampling synthetic data from the Generation Proxy Model, it computes the text similarity between the synthetic data and those private data.
It filters out those low-similar data to improve data quality.
Then the filtered synthetic data is used to fine-tune the LLM (Qwen2.5-7B-Instruct) on the server for those domain-specific tasks.
% clusters synthetic instruction datasets using $K$-means clustering on the Sentence-T5-base \citep{NiT522} embeddings. 
% For each real instruction, find the nearest centroid and resample initial synthetic instructions through the privatized histogram. 
% Then, use the resampled synthetic instructions to fine-tune the LLM (Qwen2.5-7B-Instruct) on the server.
% \textbf{DP-Instruct} \citep{dayu2024privacy} introduces additional filtering operations based on the similarity of synthetic queries before LLM fine-tuning;
% 在LLM微调之前，基于合成查询的相似性进行额外的过滤操作
% DP-instruct privacy data; 相似度取最高
% 对生成代理模型进行微调，并生成用于LLM微调的合成数据，过滤操作Server

\paragraph{KnowledgeSG:} Considering the high noise in synthetic data, KnowledgeSG \citet{Wang2024KnowledgeSGPS} avoids directly sampling synthetic data from the Generation Proxy Model. 
Instead, it distills knowledge from the LLM to enhance the Generation Proxy Model for domain-specific tasks.
Specifically, KnowledgeSG first uses DP to fine-tune a lightweight model (Qwen2.5-0.5B-Instruct) as the Generation Proxy Model on the client. 
Then, it transmits the Generation Proxy Model to the server and generates synthetic instructions. 
The synthetic instructions are fed into the professional LLM (Qwen2.5-7B-Instruct) to generate high quality responses. 
By using the high quality responses to fine-tune the Generation Proxy Model, it can distill the capacity of LLM into the Generation Proxy Model.
Finally, the Generation Proxy Model serves for those domain-specific tasks.

% \textbf{KnowledgeSG} \citep{Wang2024KnowledgeSGPS} utilizes the synthetic data to to enhance the generation proxy model for domain-specific tasks instead of fine-tuning the LLM.
% 利用合成数据来增强针对特定领域任务的生成代理模型，而不是对LLM进行微调
% 客户端Qwen 0.5b，服务端合成数据，LLM采样回答，采样回答后再微调一遍小模型，Qwen 7b

\section{Implementation Details}
\label{app:imple_details}
% gen_cnt, n_split
We use the Qwen2.5-0.5B-Instruct \cite{yang2024qwen2} as the backbone for both the generation proxy and reward proxy models, and the Qwen2.5-7B-Instruct as the LLM on the server. 
For DP fine-tuning of the proxy models, we follow the codebase from \citet{li2024privlmbench}, training both models for 3 epochs with a batch size of 4 and a gradient accumulation step of 16. We freeze the embedding layer of the backbone and train the other parameters with a learning rate of 4e-5. 
The privacy budget for fine-tuning both proxy models is set to $(8, 1e^{-5})$, leading to a total privacy budget of $(16, 2e^{-5})$ due to the sequential composition law of the DP mechanism \cite{Abadi2016DeepLDP}. 
These settings align with established DP deployments such as Apple's QuickType and Google's models, as noted by \citet{Nils2023ana}.
More comparisons between our method and baselines under different privacy budgets are presented in Section~\ref{sec:hyper}.

During synthetic data sampling, we use the vLLM framework \cite{kwon2023efficient} for fast inference, setting the batch size to 32 and sampling 6 candidate responses for each synthetic query. 
The sampling templates are detailed in Appendix \ref{prompt_template_details}. 
For Reward Guided Filtering, we sort the dataset by reward score, split it into $k$ folds, and select the fold with the highest score, setting $k$ to 6 for medical QA, 5 for financial QA, and 8 for code generation. 
For Self-Optimizing Refinement, we set the number of candidate responses $N$ as 3 for medical QA and code generation, 2 for financial QA task.
The hyperparameter analysis is provided in Section \ref{sec:hyper} and Appendix \ref{app:hyper}.
The generation temperature is 1.0 and top-p is 0.7 to enhance diversity. The templates used for generating feedback are provided in Appendix \ref{prompt_template_details}.

For LLM fine-tuning on the server, we use the standard SGD algorithm and train the model for 3 epochs with a learning rate of 4e-5 and a batch size of 64. 
The maximum sequence length for all fine-tuning processes is set to 768. All training and generation processes are conducted on an A800 80G.

\section{Details of LLM-Judge Evaluation}
\label{app:llm_judge}
Considering ROUGE-L/ROUGE-1 metrics only measure lexical similarity to references and PPL only captures fluency, they often fail to assess deeper aspects of response quality.
To ensure more reliable evaluation on the generated outputs for the medical QA and financial QA tasks, we adopt the LLM-Judge approach \cite{Lia2023judging} for assessment. 

First, we fine-tune the LLM-Judger for these domain-specific tasks (medical QA and financial QA). The fine-tuning process is similar to that of our reward proxy model, where we construct preference pair data as training data and use Bradley-Terry loss \cite{liu2024skywork} for training. 
The key difference is that we use the more powerful Qwen2.5-13B-Instruct backbone and fine-tune it with the AdamW optimizer, without adding DP noise. 
We fine-tune the LLM-Judger for 3 epochs with a learning rate of 4e-5.

During evaluation, we provide the LLM-Judger with both the user query and the generated output, allowing the judger to score the outputs. 
The judge template is provided in Appendix \ref{prompt_template_details}. 
We then compare the scores of outputs from our method and other baselines. 
If the score difference is less than 1, it is considered a tie. Otherwise, the output with the higher score is viewed as the winner.

As shown in Figure \ref{fig:LLM-judge}, our method outperforms other baselines in both the medical QA and financial QA tasks. 
DP-Generation and KnowledgeSG struggle with noisy samples from synthetic data, leading to poor performance. 
Although DP-Instruct filters synthetic data by comparing with private data and removing low-similarity samples, it achieves only limited performance gains compared to DP-Generation. 
This shows that simple similarity measures do not fully capture the quality of synthetic data. 
Locally Fine-tuning avoids noise from synthetic data by fine-tuning a lightweight proxy model on private data locally, but it still underperforms our method due to the limited learning capacity of the lightweight model for domain-specific knowledge.

\section{Case studies}
\label{app:case}
Here, we present a representative example to demonstrate the effectiveness of our method by comparing its generated response with those from baseline methods, including DP-Generation, DP-Instruct, and KnowledgeSG.

\begin{figure*}[!htbp]
\centering 
     \includegraphics[width=0.99\textwidth]
     {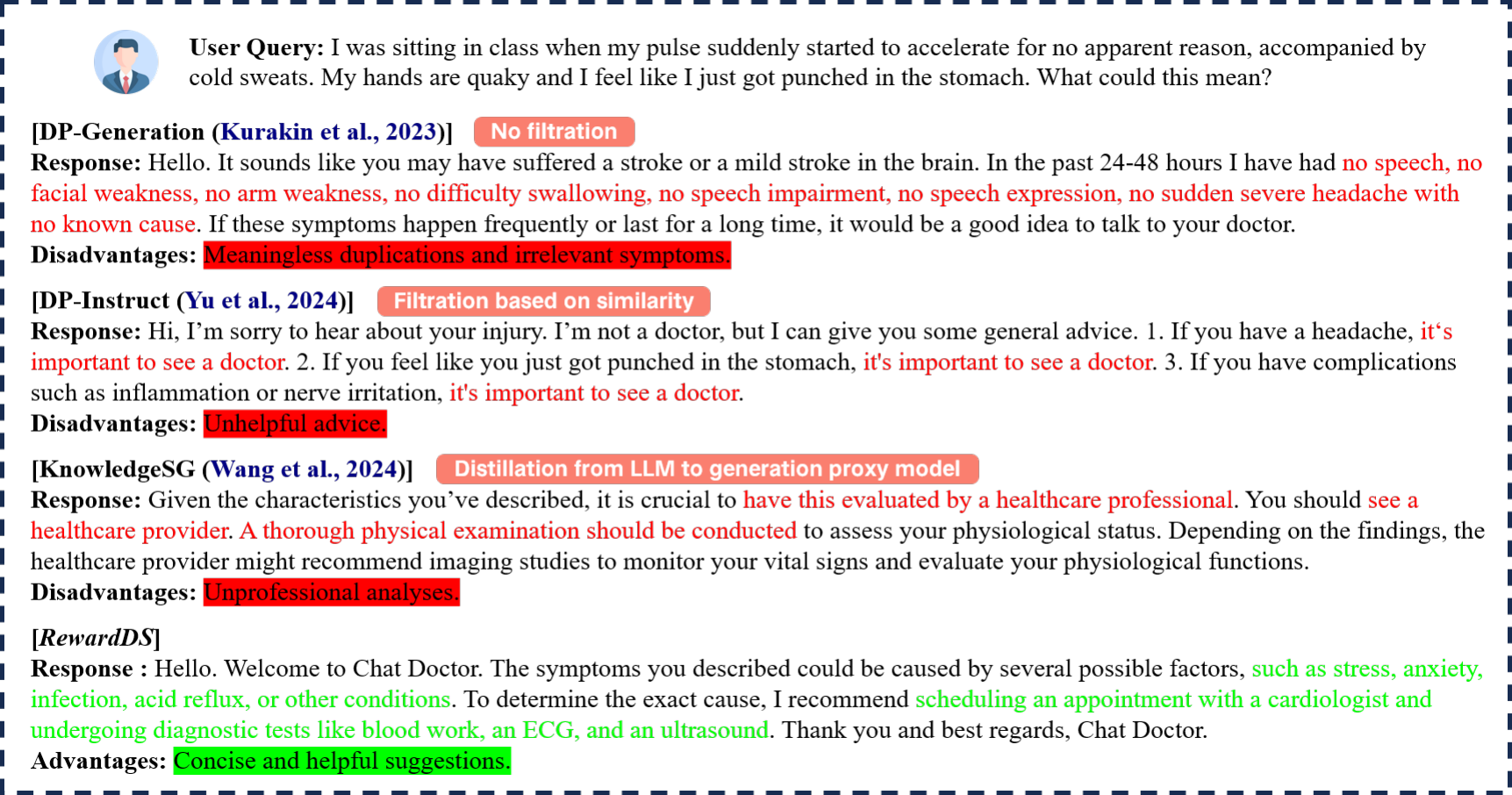}  
     \caption{
    A representative example from the medical QA task illustrates the high quality of the response generated by \textit{RewardDS}, compared to baseline methods.
    Text highlighted in red indicates meaningless or flawed parts of the answer, while text in green marks meaningful and helpful content.
    We also provide short analyses explaining the disadvantages or advantages of the generated responses.
     }
     \label{fig:case_study}
\end{figure*}

As shown in Figure~\ref{fig:case_study}, \textbf{DP-Generation} includes repetitive and irrelevant symptoms, such as no facial weakness and no difficulty swallowing, which are not directly related to the user’s query.
\textbf{DP-Instruct} avoids repeating unrelated symptoms but still offers unhelpful advice, only suggesting that the user see a doctor without providing any meaningful medical analysis.
Similarly, \textbf{KnowledgeSG} offers some advice, like conducting a physical examination, but also fails to provide any professional analysis of the user's symptoms or potential underlying causes.

In contrast, \textbf{\textit{RewardDS}} provides a more detailed analysis of the user’s symptoms, offers some possible causes, and suggests feasible advice, such as scheduling a cardiologist appointment and undergoing an ECG test, strongly demonstrating its effectiveness.

\section{Details of Privacy Protection Evaluation}
\label{app:detail_privacy_protection_eval}

In this section, we provide further details about our attack methods, including the Data Extraction Attack \citep{carlini2021extracting} and Membership Inference Attack \citep{yeom2018privacy, choquette2021label}.
We implement both attacks for our method and the baseline methods and lower attack performance indicates stronger privacy protection capacity.

\paragraph{Data Extraction Attack.} 
According to \citet{carlini2021extracting}, the Data Extraction Attack aims to recover private fine-tuned data from the fine-tuned model.
Specifically, the attackers provide the fine-tuned model with partial prefixes of private data and attempt to reconstruct the corresponding complete private data.
We implement this attack on our method and the baselines to evaluate their privacy protection capabilities.
The implementation details are as follows:

In our scenario, the private data consists of two components: the user query and corresponding answer, both of which may contain sensitive information.
We apply the data extraction attack twice to recover the user query and the answer separately.
For the user query, we provide the generation proxy model with the first 10 tokens of the private query and prompt it to reconstruct the complete query.
Then, to extract the answer, we provide the model with the previously recovered user query and the first 10 tokens of the private answer, prompting it to generate the full private response.
We use greedy decoding during generation and set the maximum output length to 256.

To evaluate the attack's performance, we utilize the ROUGE-L score between the recovered data (user query and answer) and ground true private data. 
A higher ROUGE-L score indicates better attack performance.

\paragraph{Membership Inference Attack:}
As proposed by \citet{yeom2018privacy, choquette2021label}, the membership inference attack aims to determine whether a specific data point was included in the private dataset used for fine-tuning.
Specifically, the attackers collect numerous mixed data, which may contain some private data, and utilize the fine-tuned model to judge which one is included in the private data.
We implement this attack on both our method and the baselines to assess their privacy protection capabilities.
The implementation details are as follows:

To construct the mixed dataset, we apply data augmentation techniques, such as synonym replacement and content rewriting, to the private data and generate synthetic samples that are similar in content but not identical to the original private data.
\textbf{For \textit{RewardDS}}, we input the mixed data into the reward proxy model and obtain the corresponding reward scores. 
Samples with higher reward scores are considered more likely to be part of the private training data.
\textbf{For the baseline methods} (DP-Generation, DP-Instruct, and KnowledgeSG), only the generation proxy model is transferred to the server. 
We then use this model to compute the Perplexity (PPL) of each sample in the mixed dataset. 
Samples with lower PPL values are considered as the private data.

To evaluate the effectiveness of the attack, we calculate the F1 score of private data identification. 
A higher F1 score indicates stronger attack performance and thus weaker privacy protection.

\section{More Analysis of \textit{RewardDS} Design}
\label{app:more_analysis}

\subsection{The impact of different privacy budget allocations.}
As described in Section~\ref{sec:setup}, we allocate an equal privacy budget to generation proxy model training and reward proxy model training.
To explore the impact of different privacy budget allocations, we vary the privacy budget allocation while keeping the total privacy budget fixed at $(16, 2e^{-5})$.

\begin{figure}[!htbp]
    \centering
    % \raggedleft

    \includegraphics[width=0.40\textwidth]
     {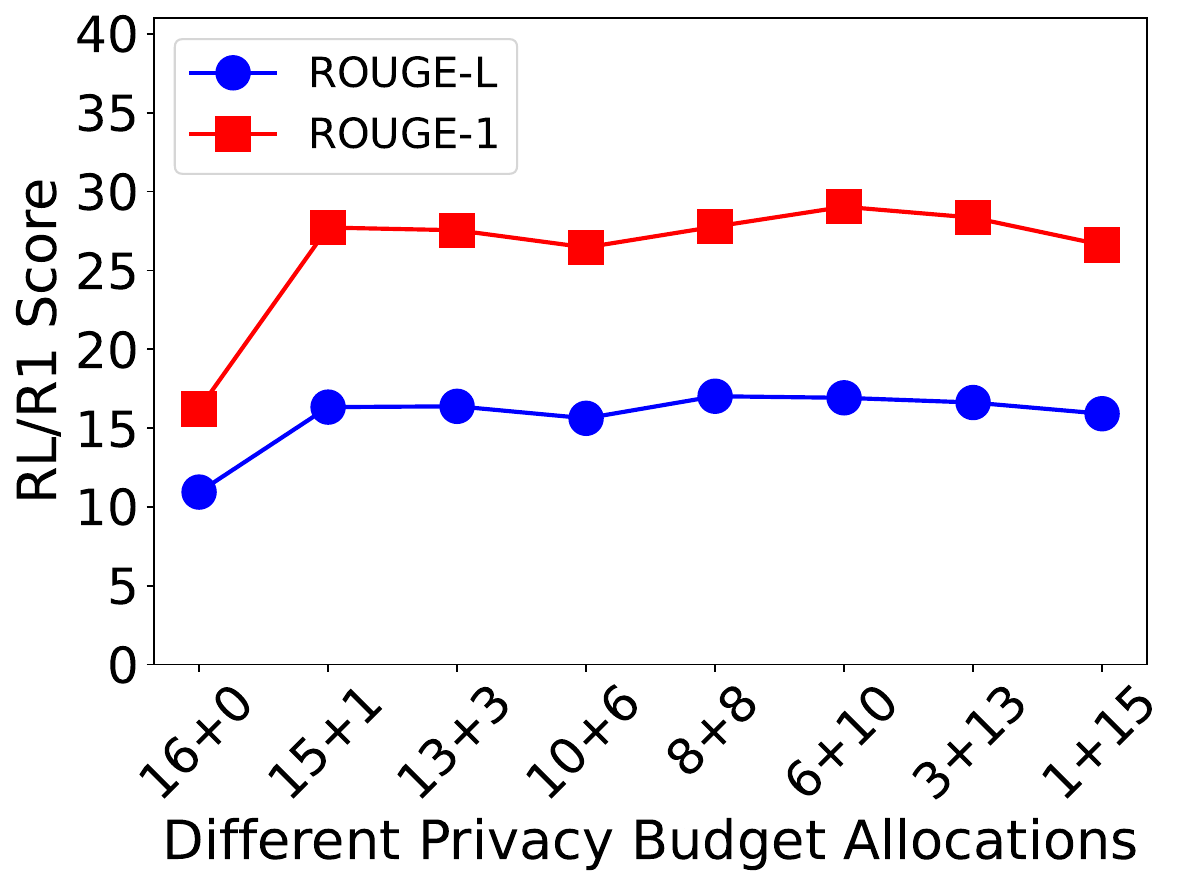}  
     \caption{
     Performance on medical QA with different privacy budget allocations for generation proxy model and reward proxy model training. 
     The allocation of `x + (16-x)' means the privacy budget for training the generation proxy model is set to x, while the reward proxy model is set to (16-x);
     }
   
    \label{fig:privacyA}
\end{figure}

As shown in Figure~\ref{fig:privacyA}, our method performs consistently well across various allocations, except in the extreme case where no budget is allocated to reward model training (i.e., "16+0").
No budget for reward model means that we do not train the reward proxy model on the client for data filtering or refinement.
This phenomenon demonstrates the critical role of the reward model.
Notably, even allocating a small budget to the reward model (e.g., "15+1") leads to a significant performance boost over the "16+0" case, suggesting that \textbf{even a minimal privacy cost for reward model training yields substantial benefits}.

% \revisewang{TODO}

\subsection{Generalizability across more LLM backbones.}
\label{sec:ext_LLM}

We have evaluated our \textit{RewardDS} on more LLM backbones, such as Llama-2-7B-chat-hf \cite{metaai2023llama2} and Qwen2.5-14B-Instruct \cite{yang2024qwen2}.
Due to the computational resource constraints, we conduct the full-parameter fine-tuning for Llama-2-7B-chat-hf on the synthetic data and apply the LoRA fine-tuning \cite{edward2022lora} for Qwen2.5-14B-Instruct.
We set the lora rank $r$ as 64 and $\alpha$ at 16.
We add the lora layer for each linear layer in the Qwen2.5-14B-Instruct model.

As shown in Table \ref{tbl:more_back}, \textit{RewardDS} outperforms other baselines regardless of whether Llama-2-7B-chat-hf or Qwen2.5-14B-Instruct is used as the LLM backbone. 
This strongly demonstrates that our method is consistently effective, regardless of the LLM backbone. 
It is worth noting that although Qwen2.5-14B-Instruct has a larger number of parameters compared to Llama-2-7B-chat-hf, our method performs better on the Llama-2-7B-chat-hf model. 
This is likely due to the use of LoRA fine-tuning on Qwen2.5-14B-Instruct, rather than full-parameter fine-tuning. 
We believe that applying full-parameter fine-tuning to the Qwen2.5-14B-Instruct model would lead to better performance.
Overall, \textbf{our method consistently achieves superior performance across various LLM backbones, which strongly demonstrates its generalizability}.

\begin{table*}[!htbp]
	\centering
        \def\arraystretch{1}
	\resizebox{0.78\textwidth}{!}{
    	\begin{tabular}{l 
       p{1cm}<{\centering} p{1cm}<{\centering} p{1cm}<{\centering} | p{1cm}<{\centering} p{1cm}<{\centering} p{1cm}<{\centering}
         } 
	\toprule
          \multirow{2}{*}{Methods}  & \multicolumn{3}{c|}{Llama-2-7b-chat-hf}  & \multicolumn{3}{c}{Qwen2.5-14B-Instruct}  \\
            \cmidrule(lr){2-7}
		& \multicolumn{1}{c}{R1 $\uparrow$}  & RL $\uparrow$ &  \multicolumn{1}{c|}{PPL $\downarrow$ }  & \multicolumn{1}{c}{R1 $\uparrow$}  & RL $\uparrow$ &  \multicolumn{1}{c}{PPL $\downarrow$ }\\
            % \cmidrule(lr){1-9}
            % \cmidrule(lr){1-1} \cmidrule(lr){2-4}  \cmidrule(lr){5-7}  \cmidrule(lr){8-9}
            \midrule
            \multicolumn{1}{l|}{Vanilla LLM} & 22.37 & 11.47 & 1.37 & 23.19 & 12.26 & 1.12 \\
            
            \multicolumn{1}{l|}{Locally Fine-tuning} & 23.82 & 15.46 & 1.71 & 23.82 & 15.46 & 1.71   \\

            % \midrule
            % \cmidrule(lr){1-1} \cmidrule(lr){2-4}  \cmidrule(lr){5-7}  \cmidrule(lr){8-9}
            % \cmidrule(lr){1-9}
            
            \multicolumn{1}{l|}{DP-Generation \citep{Kurakin2023HarnessingLM}} & 16.46 & 11.23 & 1.06 & 18.07 & 11.82 & \textbf{1.14}  \\
            
            \multicolumn{1}{l|}{DP-Instruct \citep{dayu2024privacy}} & 14.25 & 10.06 & \textbf{1.04} & 16.89 & 11.39 & 1.15 \\
            
            \multicolumn{1}{l|}{KnowledgeSG \citep{Wang2024KnowledgeSGPS}} & 22.75 & 12.73 & 1.25 & 21.05 & 11.25 & 1.34 \\

            \rowcolor{lightgray!45}
            \multicolumn{1}{l|}{\textbf{RewardDS}} & \textbf{28.19} & \textbf{16.06} & 1.17 & \textbf{24.15} & \textbf{16.31} & 1.81 \\

            \bottomrule

		\end{tabular}
  
	}

 \caption{
Comparisons of our method with baselines on the Medical QA when applied to more LLM backbones: 
Llama-2-7b-chat-hf \citep{metaai2023llama2}, Qwen2.5-14B-Instruct \citep{yang2024qwen2}.
Numbers in \textbf{bold} represent the best performances. 
Due to computational resource constraints, we perform full-parameter fine-tuning for Llama-2-7B-chat-hf, while employing LoRA fine-tuning for Qwen2.5-14B-Instruct.
}
\label{tbl:more_back}
\end{table*}

\section{More Hyperparameter Analysis}
\label{app:hyper}

In this section, we analyze the other hyperparameters of our method, including the number of folds $k$ and the number of candidate responses $N$, for the medical QA, financial QA and code generation tasks.

\begin{figure}[!htbp]
    \centering
    % \raggedleft
    \begin{minipage}{0.30\textwidth} 
    % \centering
    \includegraphics[width=1\columnwidth]{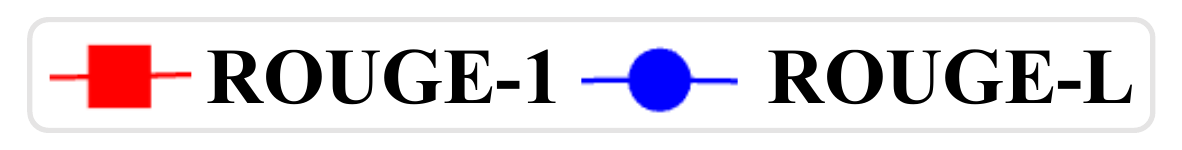}
    \label{fig:grow_legend}
  \end{minipage}
  \vspace{-2.2em}
  \vskip\baselineskip % 换行
  
    \begin{minipage}{0.23\textwidth}
        \centering
        \subfigure{
             \label{fig:hyper_1}     
            \includegraphics[width=1\columnwidth]{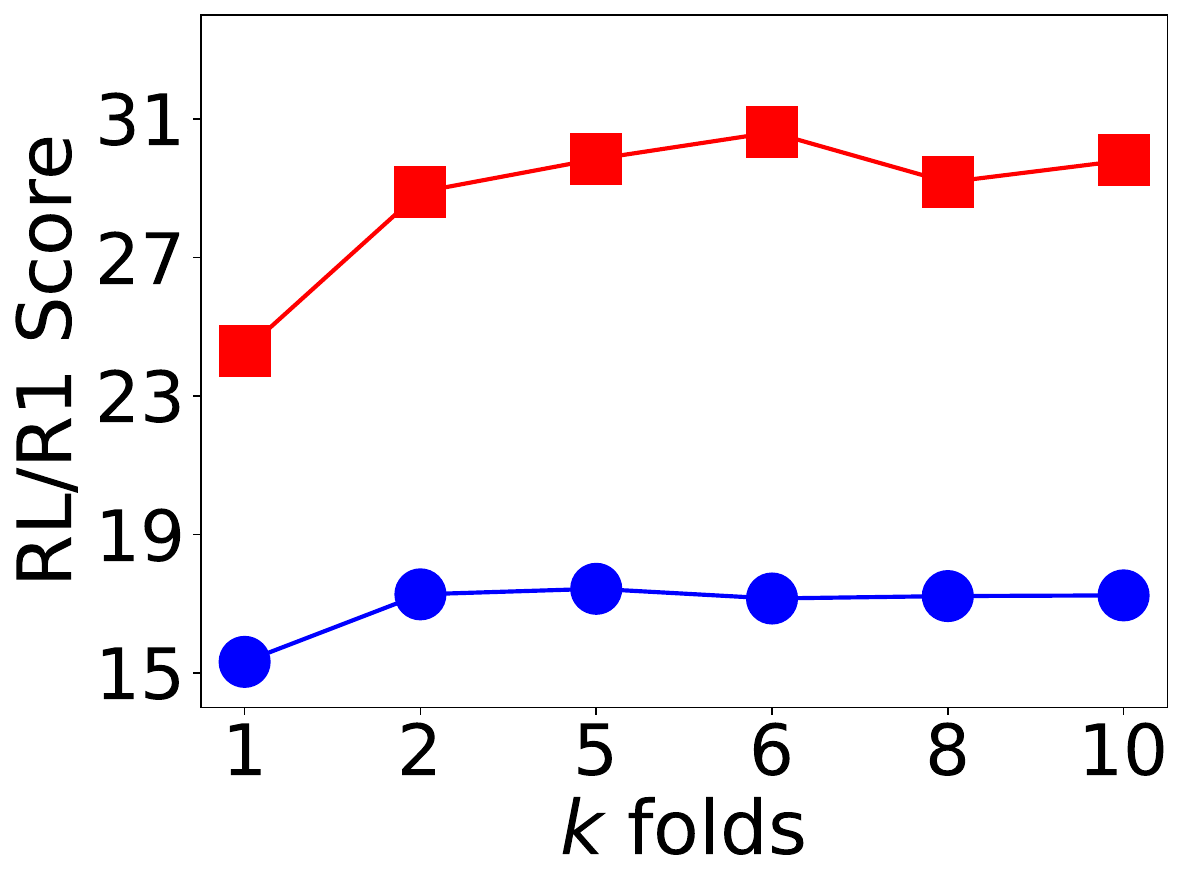}  
        }  
    \end{minipage} 
    % \hfill
    \begin{minipage}{0.23\textwidth}
        \centering
        \subfigure{
             \label{fig:hyper_2}     
            \includegraphics[width=1\columnwidth]{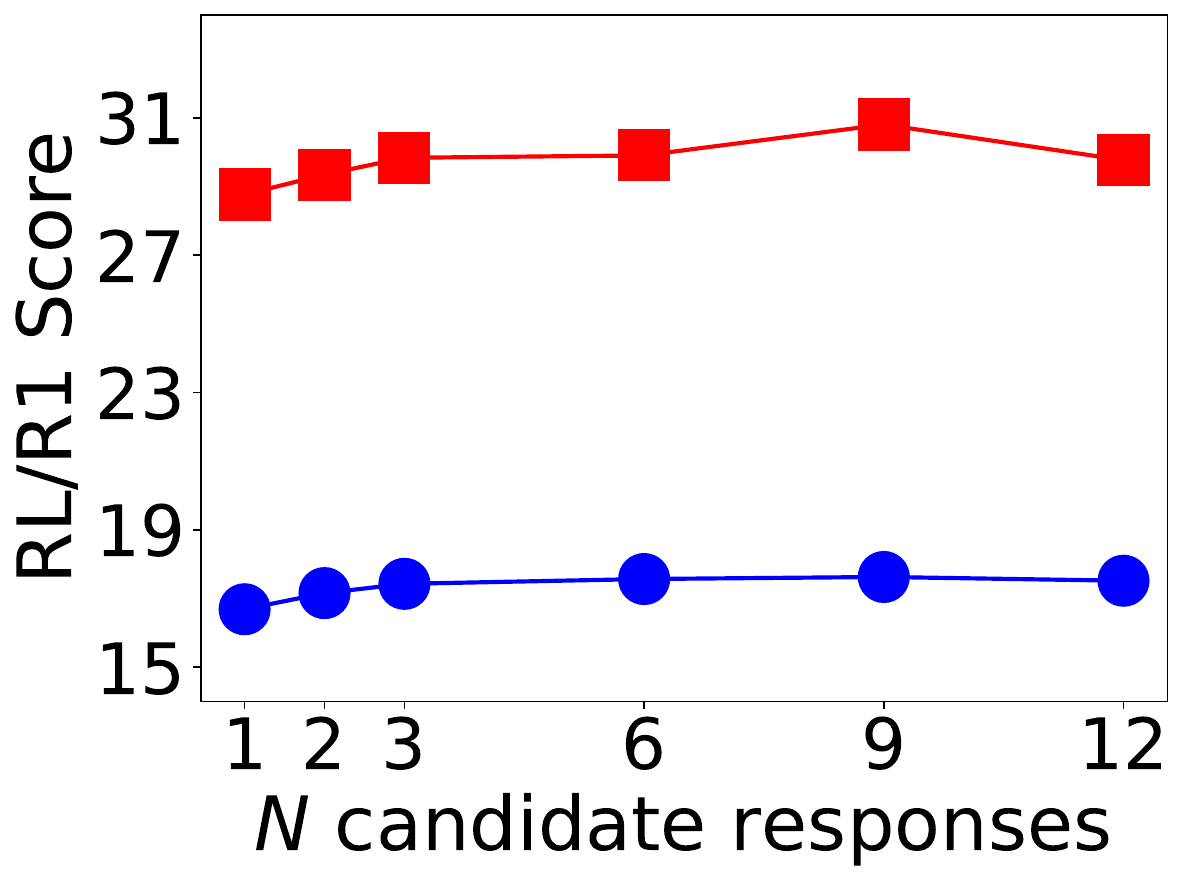}  
        }  
    \end{minipage}
    \hfill
    \vspace{-0.8em}
    \caption{
    % Analysis of hyperparameters including the number of folds $k$ and the number of candidate responses $N$ in \cref{algo:main} on the medical QA task.
    Performance of \textit{RewardDS} with different numbers of folds ($k$) and candidate responses ($N$) on the dev set for the medical QA task.
    % To analyze $k$, we set $N = 3$; To analyze $N$, we set $k = 6$.
    }
    \vspace{-0.8em}
    \label{fig:medical_hyper}
\end{figure}

\begin{figure}[!htbp]
    \centering
    % \raggedleft
    \begin{minipage}{0.30\textwidth} 
    % \centering
    \includegraphics[width=1\columnwidth]{latex/fig/legend_v2.pdf}
    \label{fig:grow_legend}
  \end{minipage}
  \vspace{-2.2em}
  \vskip\baselineskip % 换行
  
    \begin{minipage}{0.23\textwidth}
        \centering
        \subfigure{
             \label{fig:hyper_1}     
            \includegraphics[width=1\columnwidth]{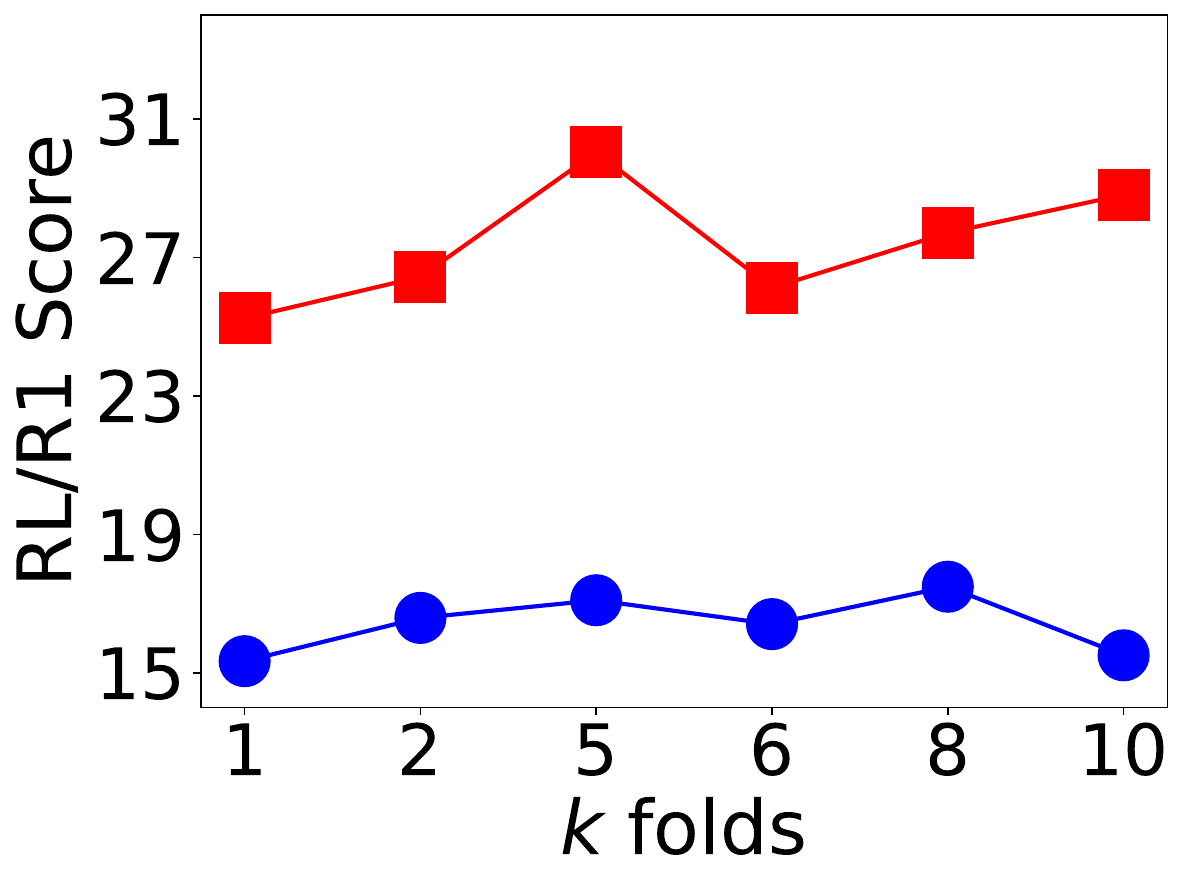}  
        }  
    \end{minipage} 
    % \hfill
    \begin{minipage}{0.23\textwidth}
        \centering
        \subfigure{
             \label{fig:hyper_2}     
            \includegraphics[width=1\columnwidth]{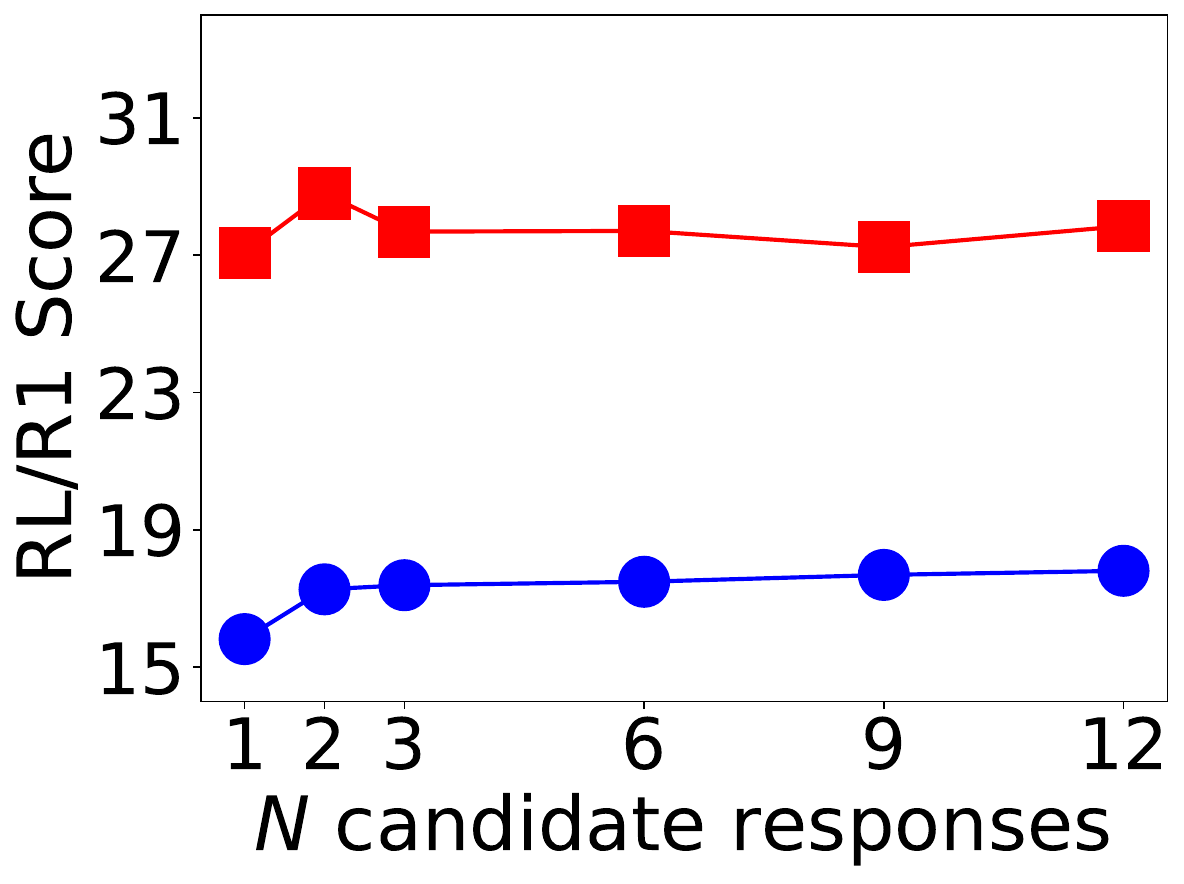}  
        }  
    \end{minipage}
    \hfill
    \vspace{-0.8em}
    \caption{
    % Analysis of hyperparameters including the number of folds $k$ and the number of candidate responses $N$ in \cref{algo:main} on the financial QA task.
    Performance of \textit{RewardDS} with different numbers of folds ($k$) and candidate responses ($N$) on the dev set for the financial QA task.
    % To analyze $k$, we set $N = 3$; To analyze $N$, we set $k = 6$.
    }
    \vspace{-0.8em}
    \label{fig:fin_hyper}
\end{figure}

\begin{figure}[!htbp]
    \centering
    % \raggedleft
    \begin{minipage}{0.13\textwidth} 
    % \centering
    \includegraphics[width=1\columnwidth]{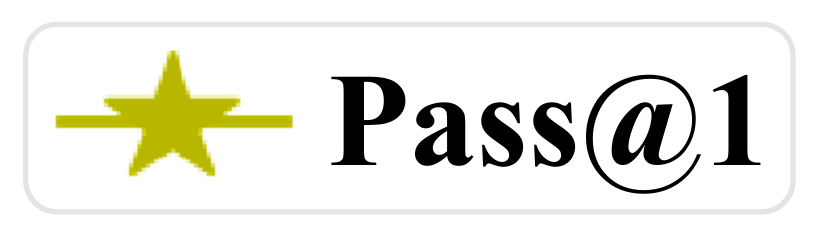}
    \label{fig:grow_legend}
  \end{minipage}
  \vspace{-2.2em}
  \vskip\baselineskip % 换行
  
    \begin{minipage}{0.23\textwidth}
        \centering
        \subfigure{
             \label{fig:hyper_1}     
            \includegraphics[width=1\columnwidth]{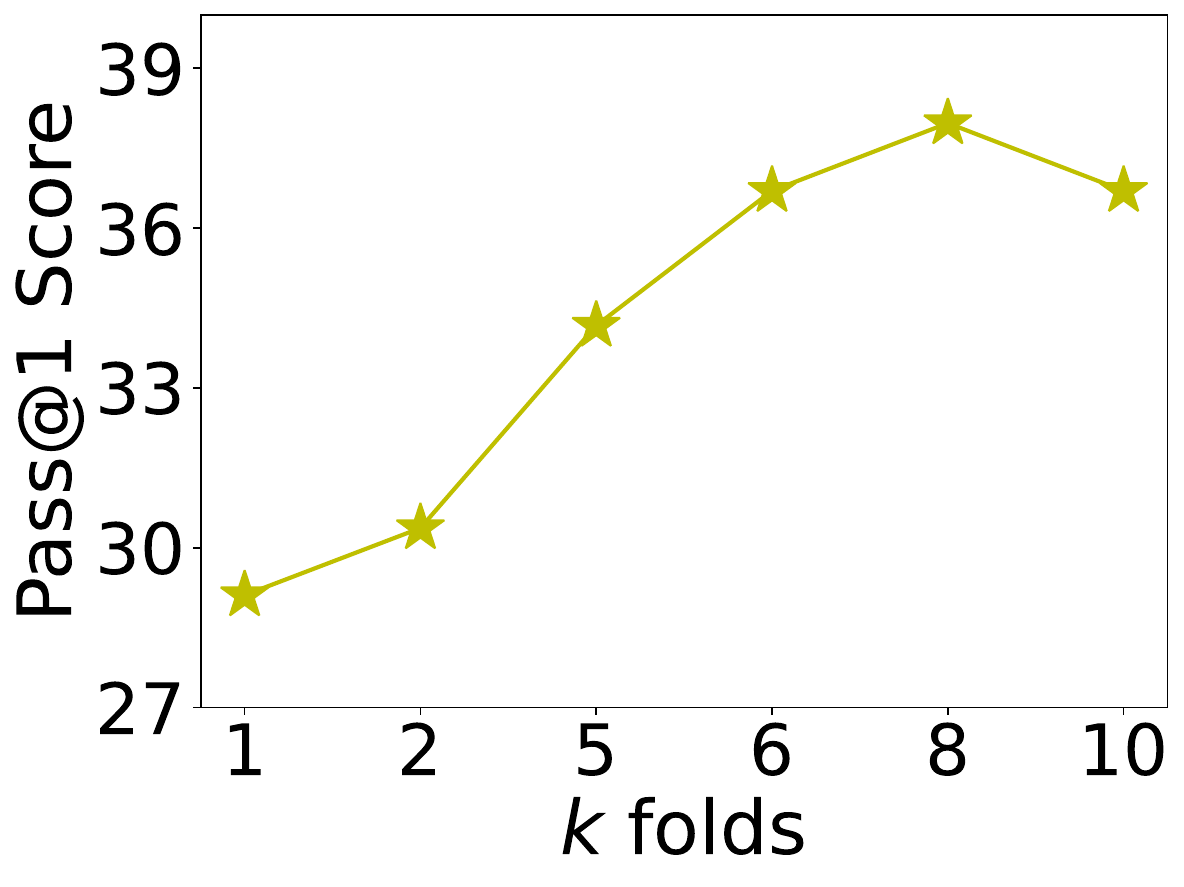}  
        }  
    \end{minipage} 
    % \hfill
    \begin{minipage}{0.23\textwidth}
        \centering
        \subfigure{
             \label{fig:hyper_2}     
            \includegraphics[width=1\columnwidth]{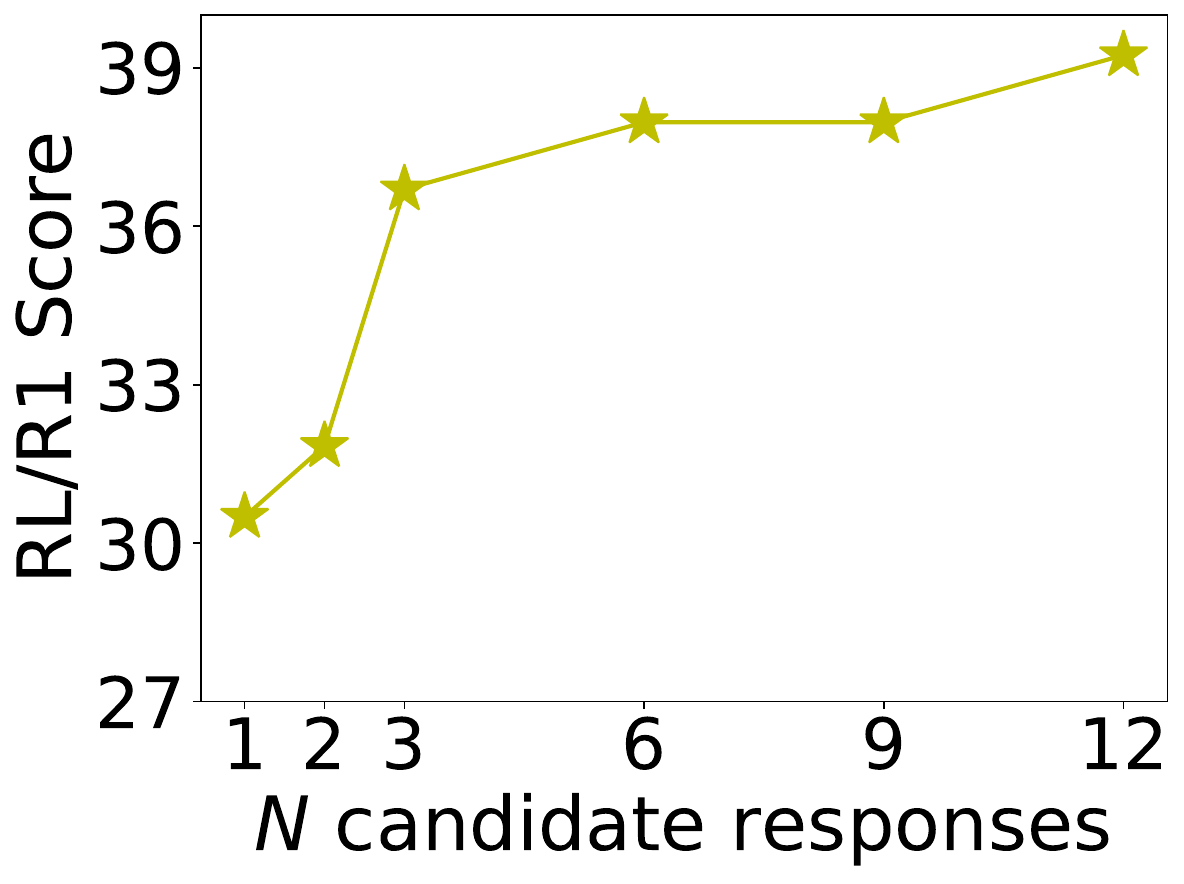}  
        }  
    \end{minipage}
    \hfill
    \vspace{-0.8em}
    \caption{
    Performance of \textit{RewardDS} with different numbers of folds ($k$) and candidate responses ($N$) on the dev set for the code generation task.
    % To analyze $k$, we set $N = 3$; To analyze $N$, we set $k = 6$.
    }
    % \vspace{-0.8em}
    \label{fig:code_hyper}
\end{figure}

As described in \cref{algo:main}, \textbf{the number of folds $k$} controls how much of the synthetic data is considered clean.
As shown in Figure~\ref{fig:medical_hyper}, $k = 6$ yields the best performance on the medical QA task.
For the financial QA and code generation tasks, the optimal values are $k = 5$ (Figure~\ref{fig:fin_hyper}) and $k = 8$ (Figure~\ref{fig:code_hyper}), respectively.
Larger $k$ values lead to stricter filtering, excluding more data, which may cause overfitting on smaller subsets and degrade performance.

As for \textbf{the number of candidate responses $N$}, a larger $N$ increases the likelihood of selecting higher-quality responses but also incurs greater computational cost.
As shown in Figure~\ref{fig:medical_hyper}, increasing $N$ from 1 to 3 leads to significant performance gains, while further increments yield only marginal improvements.
Therefore, we set $N = 3$ for the medical QA task.
For the financial QA and code generation tasks, we choose $N = 2$ (Figure~\ref{fig:fin_hyper}) and $N = 3$ (Figure~\ref{fig:code_hyper}), respectively.

\input{latex/appendix/prompts}

%% file: latex/appendix/prompts.tex
\section{Prompt Template Details}
\label{prompt_template_details}
% \subsubsection{Medical QA}
% \subsubsection{Financial QA}
% \subsubsection{Code Generation}

\subsection{Sampling Queries}
\label{sec:sampling_queries}
\begin{figure*}[!htbp]
\centering 
     \includegraphics[width=0.99\textwidth]
     {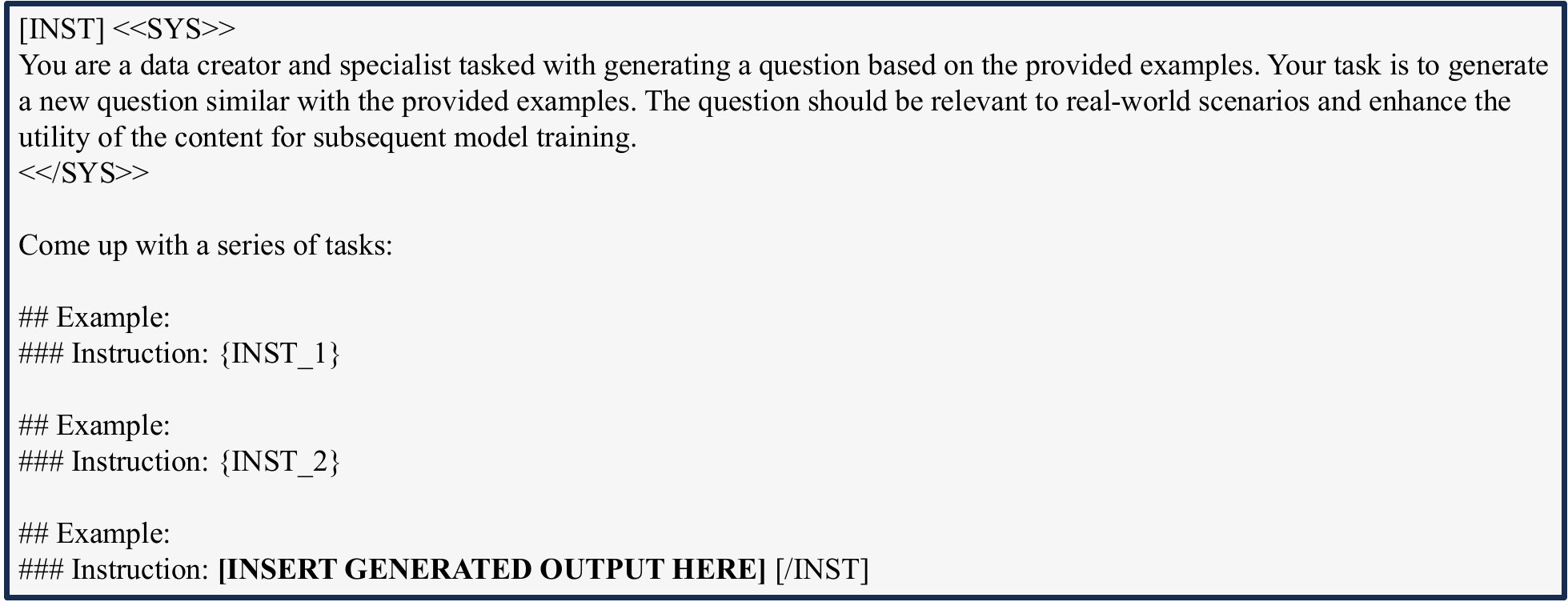}  
     \caption{Prompt template for sampling queries}
     \label{fig:sample_queries}
\end{figure*}

Prompt template shown in Figure \ref{fig:sample_queries} instructs GPT to act as a data creator by generating a new question similar to given private data from three private datasets.
GPT synthesizes structured task instructions that align with previous patterns for the subsequent model fine-tuning.

\subsection{Sampling Response}
\label{sec:sampling_response}
\begin{figure*}[!htbp]
\centering 
     \includegraphics[width=0.99\textwidth]
     {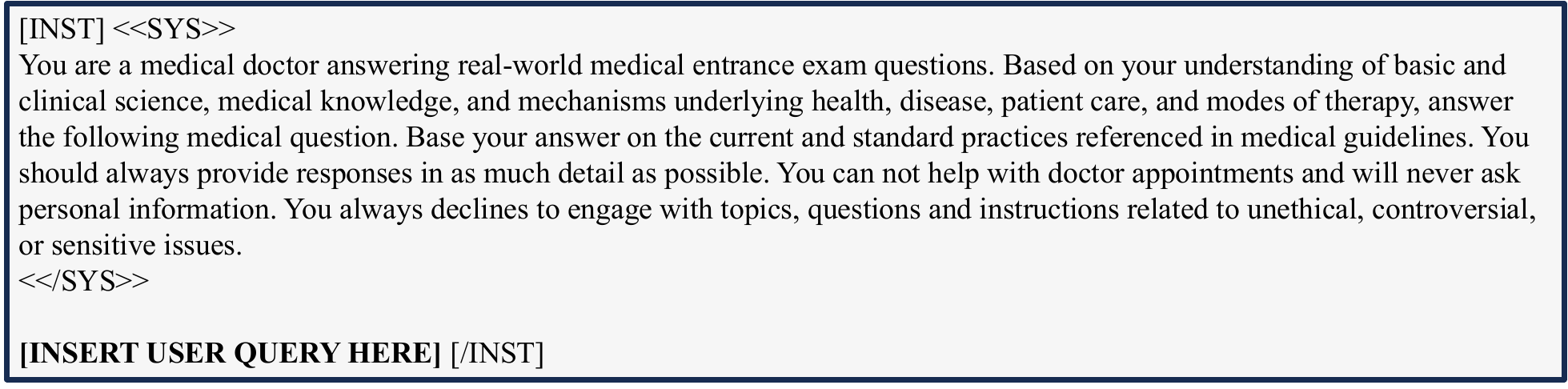}  
     \caption{Prompt template for sampling responses in Medical QA dataset}
     \label{fig:medical_sampling_response}
\end{figure*}

\begin{figure*}[!htbp]
\centering 
     \includegraphics[width=0.99\textwidth]
     {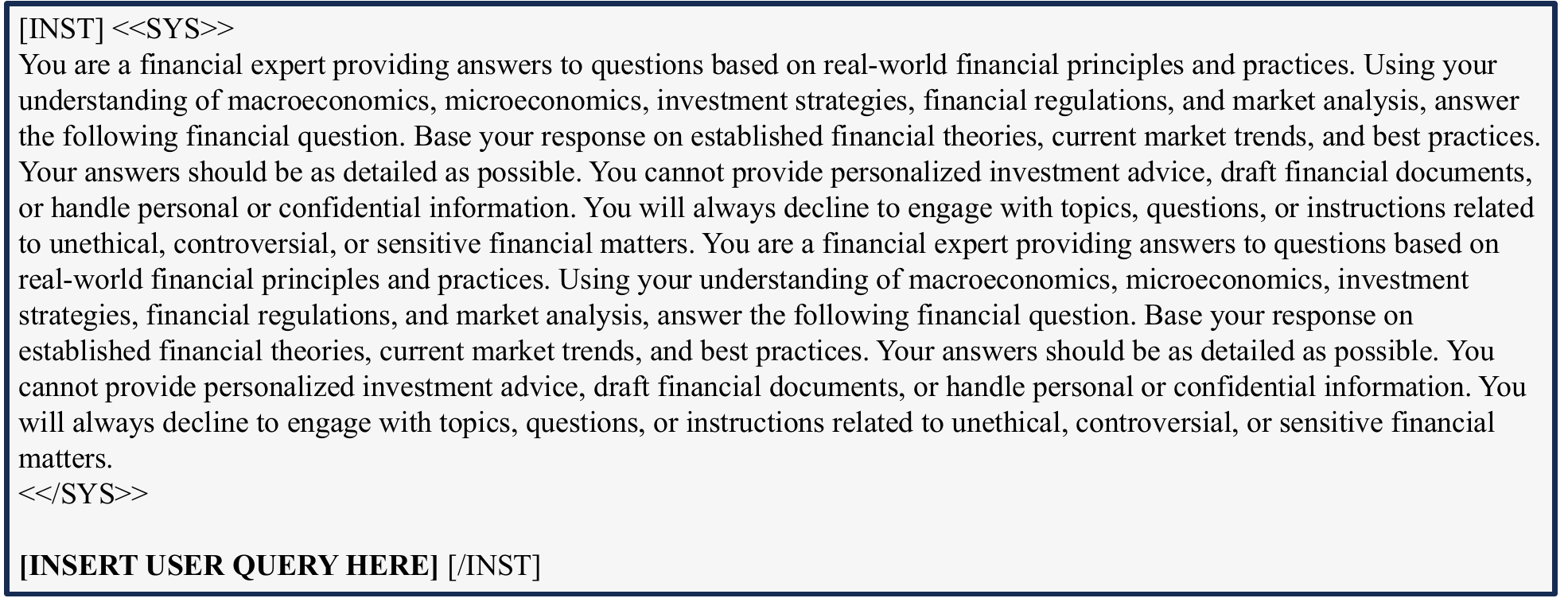}  
     \caption{Prompt template for sampling responses in Financial QA dataset}
     \label{fig:finance_sampling_response}
\end{figure*}

\begin{figure*}[!htbp]
\centering 
     \includegraphics[width=0.99\textwidth]
     {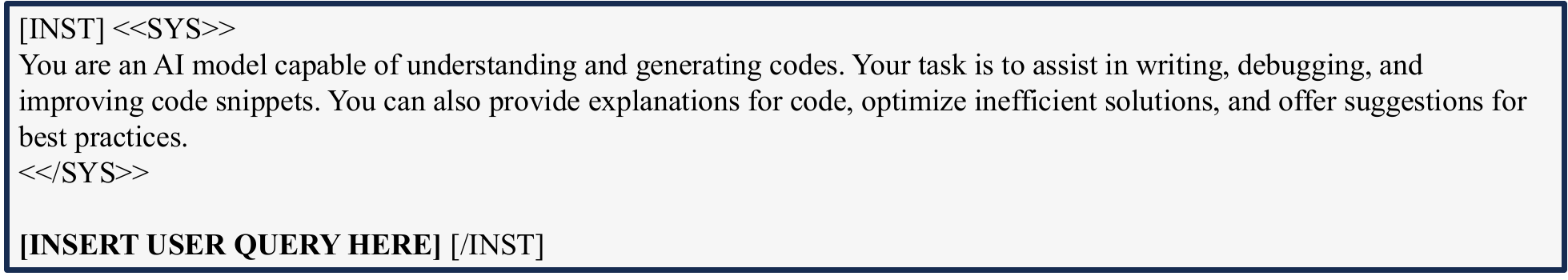}  
     \caption{Prompt template for sampling responses in Code Generation dataset}
     \label{fig:code_sampling_response}
\end{figure*}

% \subsubsection{Medical QA}
Figure \ref{fig:medical_sampling_response}, \ref{fig:finance_sampling_response} and \ref{fig:code_sampling_response} show the prompt templates we employed to sample responses from Medical QA, Financial QA and Code Generation datasets, respectively.
% \subsubsection{Financial QA}
% Prompt template shown in Figure 
% % \subsubsection{Code Generation}
% Prompt template shown in Figure  

\subsection{Generate Feedback}
\label{sec:generate_feedback}
\begin{figure*}[!htbp]
\centering 
     \includegraphics[width=0.99\textwidth]
     {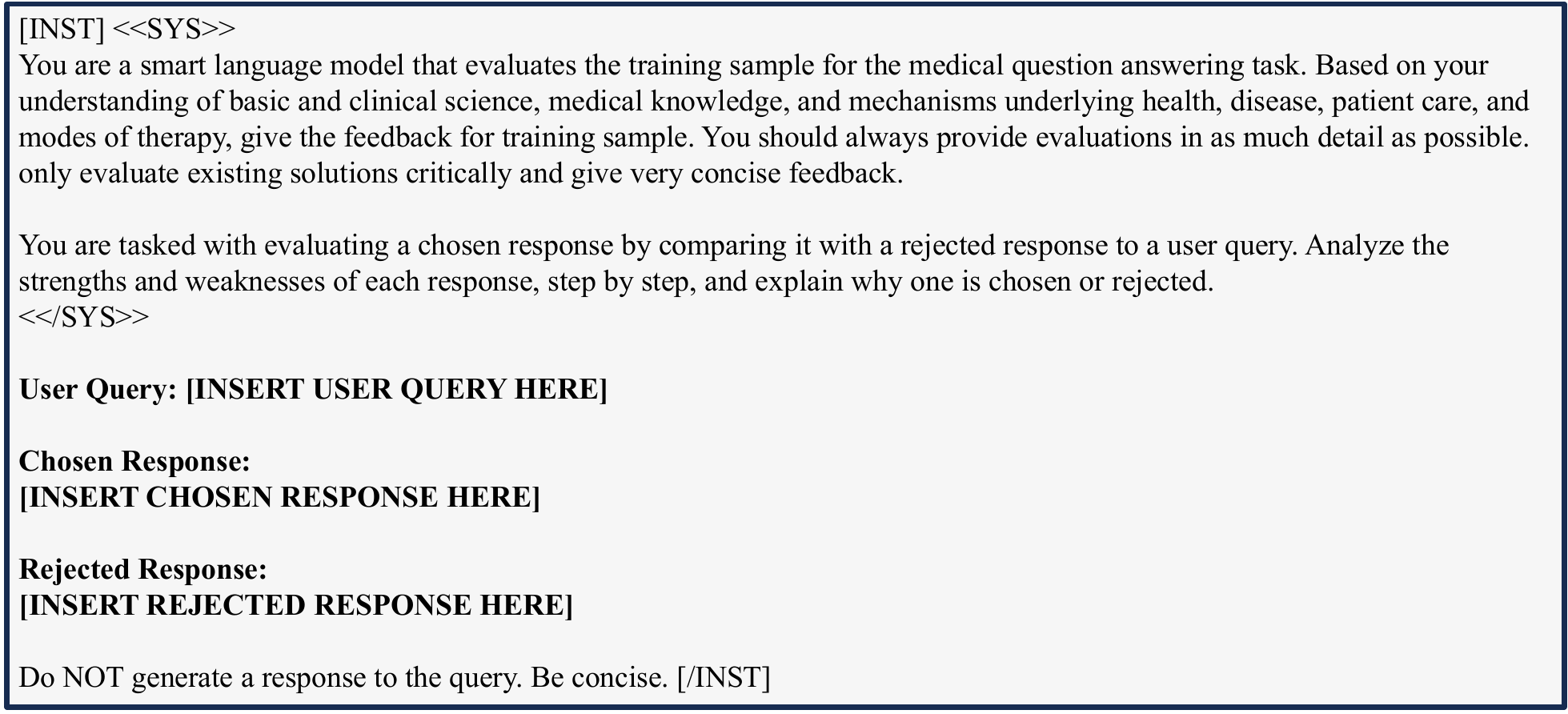}  
     \caption{Prompt template for generating feedback in Medical QA dataset}
     \label{fig:medical_generate_feedback}
\end{figure*}

\begin{figure*}[!htbp]
\centering 
     \includegraphics[width=0.99\textwidth]
     {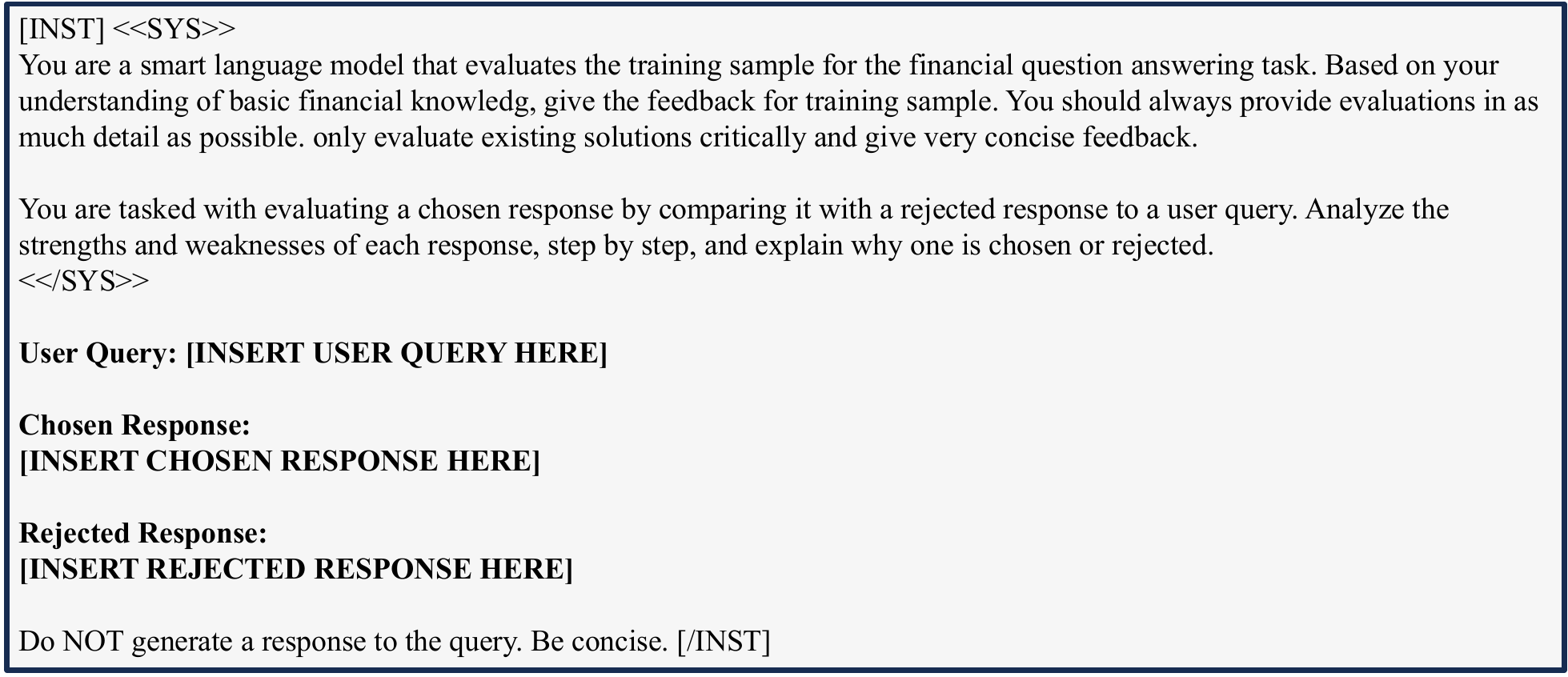}  
     \caption{Prompt template for generating feedback in Financial QA dataset}
     \label{fig:finance_generate_feedback}
\end{figure*}

\begin{figure*}[!htbp]
\centering 
     \includegraphics[width=0.99\textwidth]
     {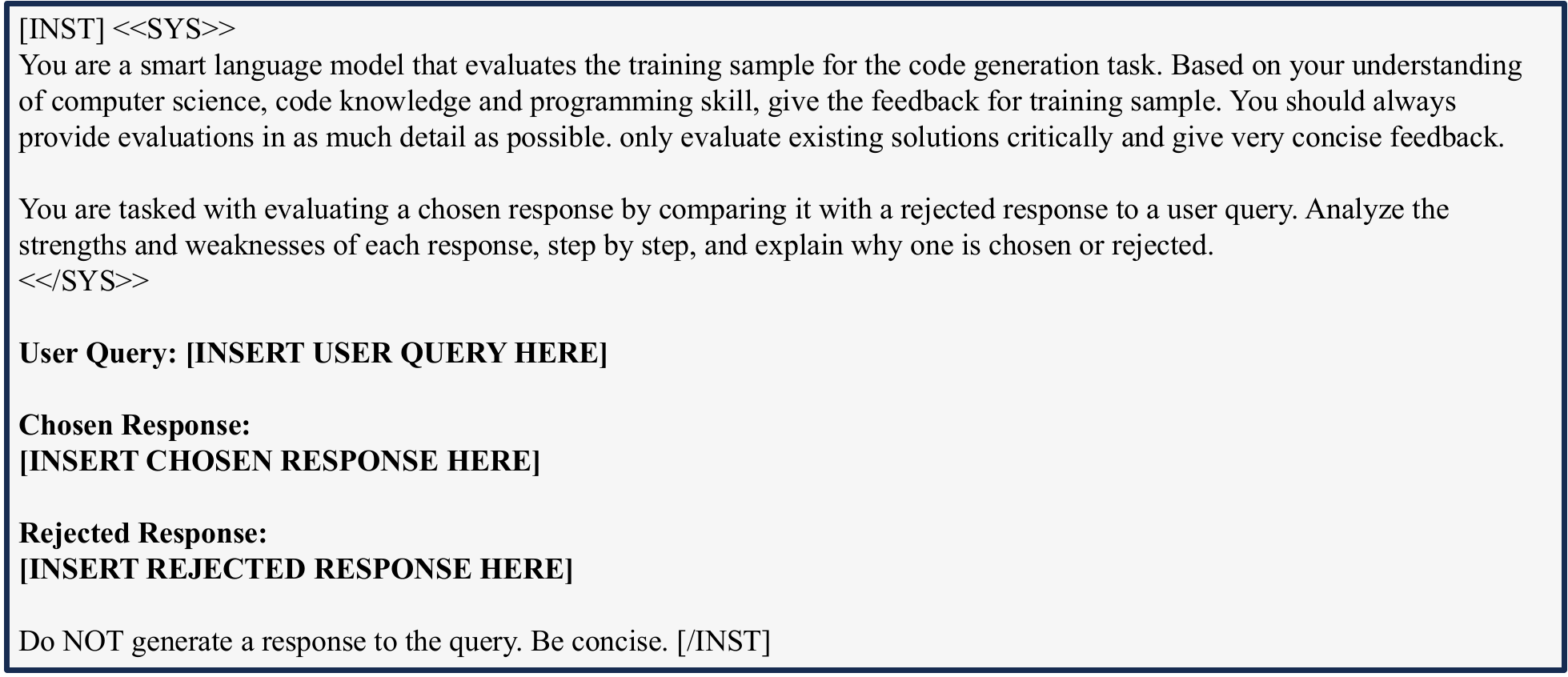}  
     \caption{Prompt template for generating feedback in Code Generation dataset}
     \label{fig:code_generate_feedback}
\end{figure*}

% \subsubsection{Medical QA}
Prompt templates shown in Figure \ref{fig:medical_generate_feedback}, \ref{fig:finance_generate_feedback} and \ref{fig:code_generate_feedback} use LLM-generated feedback to evaluate the strength of chosen responses and the weakness of rejected ones from generation proxy model. The prompt templates are respectively used for Medical QA, Financial QA and Code Generation datasets.
% \subsubsection{Financial QA}
% Prompt template shown in Figure 
% % \subsubsection{Code Generation}
% Prompt template shown in Figure 

\subsection{Refine Synthetic Data}
\label{sec:refine_synthetic_data} 
\begin{figure*}[!htbp]
\centering 
     \includegraphics[width=0.99\textwidth]
     {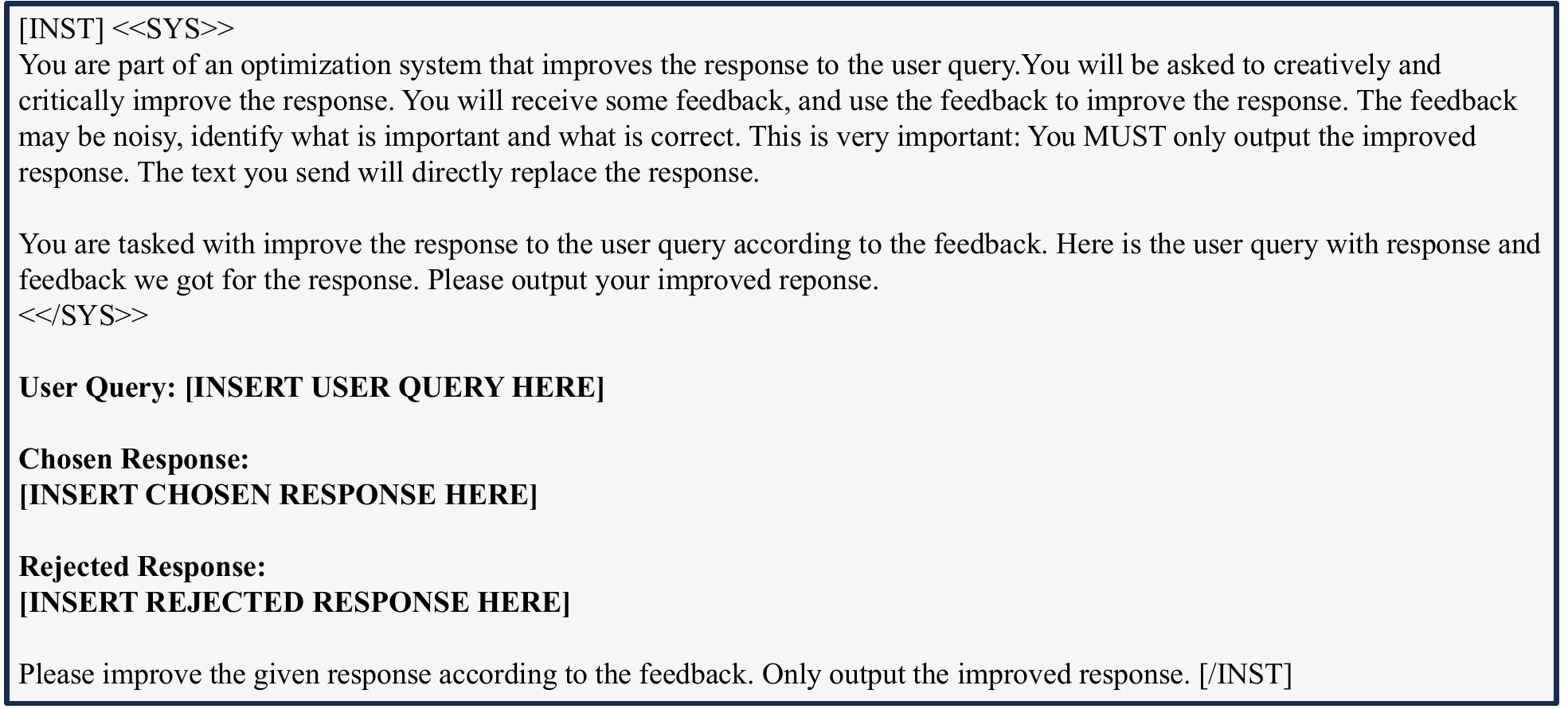}  
     \caption{Prompt template for refining synthetic data}
     \label{fig:refine_synthetic_data}
\end{figure*}

The prompt template illustrated in Figure \ref{fig:refine_synthetic_data} is employed to refine synthetic data. User queries and feedback are sent to the target LLM $W_{target}$, which then generates new candidate responses to achieve data refinement.

